\font\tenmsy=msbm10
\font\sevenmsy=msbm10 at 7pt
\font\fivemsy=msbm10 at 5pt
\newfam\msyfam 
\textfont\msyfam=\tenmsy
\scriptfont\msyfam=\sevenmsy
\scriptscriptfont\msyfam=\fivemsy
\def\blackB{\fam\msyfam\tenmsy}
\newcommand{\nc}{\newcommand}
\def\ZZ{{\blackB Z}}
\def\QQ{{\blackB Q}}
\def\RR{{\blackB R}}

\def\NN{{\blackB N}}

\def\la{{\lambda}}
\def\s{{\sigma}}
\def\st{{\tilde{\sigma}}}
\def\R{{\rangle}}
\def\L{{\langle}}
\def\d{{\partial}}

\def\be{\begin{equation}}
\def\ee{\end{equation}}
\def\su{\widehat{su}}
\def\bg{\beta\gamma}
\def\ex{\eta\xi}

\nc{\bra}[1]{\langle {#1}|}
\nc{\ket}[1]{|{#1}\rangle}

\nc{\nn}{\nonumber \\ }
\nc{\al}{\alpha}
\nc{\g}{\gamma}
\nc{\G}{\Gamma}
\nc{\D}{\Delta}
\nc{\eps}{\epsilon}
\nc{\La}{\Lambda}
\nc{\var}{\varphi}
\nc{\pa}{\partial}
\nc{\hf}{\frac{1}{2}}
\nc{\dz}{\frac{dz}{2\pi i}}
\nc{\bin}[2]{\left (\begin{array}{c} {#1}\\ {#2} \end{array}\right )}
\nc{\ben}{\begin{equation}}
\nc{\een}{\end{equation}}
\nc{\bea}{\begin{eqnarray}}
\nc{\eea}{\end{eqnarray}}

\documentstyle[epsfig]{article}
\begin{document}

\title{The $\su(2)_{-1/2}$ WZW model and the $\beta\gamma$ system}
\author{F. Lesage$^a$\footnote{lesage@crm.umontreal.ca}\ ,
P. Mathieu$^b$\footnote{pmathieu@phy.ulaval.ca}\ ,
J. Rasmussen$^a$\footnote{rasmusse@crm.umontreal.ca}\ \ and
H. Saleur$^c$\footnote{saleur@usc.edu}}
\date{}
\maketitle

\centerline{${}^a$Centre de Recherches Math\'ematiques,
    Universit\'e de Montr\'eal,}
\centerline{C.P. 6128, succursale centre-ville, Montr\'eal,
Qu\'ebec, Canada H3C 3J7}
\centerline{${}^b$D\'epartement de Physique, Universit\'e Laval,}
\centerline{Qu\'ebec, Canada G1K 7P4}
\centerline{${}^c$Department of Physics, University of Southern California,}
\centerline{Los Angeles, CA 90089-0484, USA}

\begin{abstract}
The bosonic $\beta\gamma$ ghost
system has long been used in formal constructions of conformal field 
theory. It has become  important  in its own right 
in the last few years, as a building block of field 
theory approaches to disordered systems, and as a simple 
representative -- due in part to its underlying
$\widehat{su}(2)_{-1/2}$ structure -- of non-unitary conformal
field theories. 
We provide in this paper the first complete, physical, analysis of 
this $\beta\gamma$ system, and uncover a number of striking features.
We  show in particular that the  spectrum involves  an
infinite number of fields with arbitrarily large negative dimensions. These  
fields have their origin in a twisted sector of the theory, 
and have a direct relationship
with spectrally flowed representations in the underlying
$\widehat{su}(2)_{-1/2}$ theory. We discuss the spectral flow 
in the context of the operator algebra and fusion rules, and provide
a re-interpretation of the modular invariant consistent with the spectrum.

\end{abstract}

\section{Introduction}

This is the first of two papers dedicated to the study of various
facets of $c=-1$ conformal field theory (CFT). Our general goal is to explore
in physical terms certain non-unitary  CFTs of physical importance. 
Belonging to this family of theories are the various supergroup 
Wess-Zumino-Witten (WZW) and
sigma models used in the description 
of phase transitions in disordered electronic materials
\cite{MCW,Z,Bernard,GLL,Bhaseen,SW}, the
Liouville theory which describes the conformal mode of 2D gravity
\cite{Gawedzki,Teschner}, and the
$\widehat{sl}(2,\RR)$ WZW model which plays a crucial role in the study of
strings on $AdS_3$ \cite{MO}.

A common manifestation of non-unitarity in a CFT
is the presence of  operators with a negative dimension.
The simplest example of a non-unitary CFT is the minimal $M_{2,5}$ or
Yang-Lee model, which differs little from minimal unitary CFTs. In
particular, the theory is rational and has a spectrum of dimensions
bounded from below. The models we are interested in have richer structures.
They may involve logarithms \cite{RS,G}, for example, and/or
exhibit a large, possibly infinite, set of operators with negative
dimensions.

A great deal of effort has been made to characterize and
classify abstractly such theories \cite{Flohr,Gaberdiel}. It is fair
to say, however, that very few explicit examples are well
understood. The case of $c=-2$ has given rise to
surprisingly complicated results (see \cite{Flohr,Gaberdiel} for a
review), while for potentially more interesting physical theories (such
as sigma models on superprojective spaces), partial results reveal a
truly baffling complexity \cite{ReadS}.

Our goal in this paper is to discuss another example that shares many of
the features of the more interesting models, but can still be studied in
depth. It is the $\beta\gamma$ system, which plays a crucial role in the
free-field representation of supergroup WZW models \cite{Bernard, Goddard},
for example. The $\beta\gamma$ system is a deceptively simple,
``free'' theory, with action
\bea
 S = {1\over 2\pi}\int d^2 z \left(
  \gamma_{L}\bar{\partial}\beta_{L}-\beta_{L}\bar{\partial}\gamma_{L}+
  \left(L\rightarrow R\right)\right)\label{action}
\label{betaaction}
\eea
and central charge $c=-1$. It bears a striking formal resemblance to
a non-compactified complex boson. It turns out, however, that
this model has a lot of structure, including strong non-unitary problems due
to the ill-defined nature of the functional integral (\ref{action}).
Note that the action (\ref{action})
has an obvious $Sp(2)_{R}\otimes Sp(2)_{L}$ global symmetry.

Here we will rather exploit the underlying ${su}(2)$ symmetry
discovered in \cite{LudwigNPB519}. The $\beta\gamma$ system can
actually be understood as a $\widehat{su}(2)_{-1/2}$ WZW model, as we
show here. As is well-known, the action of a WZW model based on a compact
Lie group is well-defined only when the level is integer.
An option for bypassing this obstruction for non-integer level is
to consider a non-unitary
model as being defined purely algebraically, in terms of an affine
Lie algebra at fractional
level and its representation theory.\footnote{Another option to
bypass this difficulty
would be to look for a  non-compact symmetry group, in
which case the constraint $k\in \NN$ is no longer needed. Here, we would then
consider a model with global $SL(2,\RR)$ invariance, whose spectrum
generating algebra, when $c=-1$, is $\widehat{sl}(2,\RR)_{1/2}$.}
The cornerstone of this idea is an observation by Kac and
Wakimoto \cite{KK} on $\widehat{su}(2)_k$ for fractional level
$k=t/u$ with $t\in\ZZ$ and $u\in\NN$ co-prime, and $t+2u-2\geq0$.
They found that there is a finite number of primary fields associated to
highest-weight
representations that transform linearly among themselves under the modular
group. They are called admissible representations. An example is provided
when $k=-1/2$ (in which case $c={3k}/({k+2})=-1$), inviting one to
have  the reasonable expectation that the $\widehat{su}(2)_{-1/2}$ model
is a  rational CFT.

Although the $\beta\gamma$ ghosts and twists are described  naturally
in terms of the admissible representations, we show in this paper
that neither the $\beta\gamma$ system nor the $\su(2)_{-1/2}$ WZW model is
a rational CFT in the conventional sense. In brief, this is established as
follows. Using a free-field representation of the $\beta\gamma$
system, we can show that multiple fusions of twist fields with
themselves can generate fields with arbitrarily large negative dimensions.
These are interpreted physically in terms of deeper twists. Within
the context of the WZW model, the presence of an unbounded spectrum
is explained naturally in terms of the concept of spectral flow.
A posteriori, it is then quite understandable that the non-rationality of
the WZW model at fractional level reveals itself in the context of
the hitherto puzzling issue of fusion rules. Proper
interpretations of the known fusion rules and their limitations are 
also provided.

The paper is organized as follows. In Section 2, we  review
basic facts on the $\bg$ system. We discuss thoroughly the twist
fields, and show that their $u(1)$ charge is a free parameter.
Their conformal dimension, on the other hand, is determined completely
by monodromy considerations, and is found to be $h=-1/8$.
We also discuss the
free-field representation of the $\beta\gamma$ system, which is based on
a $c=-2$ fermionic $\eta\xi$ system, and a free boson, $\phi$, with
negative metric.

Section 3 collects some results on the $\su(2)_{-1/2}$
model, such as a characterization of the admissible representations and a
description of the associated primary fields. An alternative derivation of
the $\su(2)_{-1/2}$ spectrum using the vacuum null-vector is also presented.
For later reference, we list the fusion rules computed by enforcing the
decoupling of the singular vectors \cite{AY}, as well as those
obtained by the Verlinde formula.

In Section 4, we study the $\su(2)$ structure
of the $\beta\gamma$ system, and show how the twist fields can be
organized in representations of the base $su(2)$ algebra.
We  show that a particular set of twist fields (which corresponds to
choosing a particular normal ordering for the $\beta\gamma$ system in
the Ramond sector) fit into lowest- and highest-weight
representations of spin $j=-1/4$ or $j=-3/4$.
Thus, an infinite number of twists are re-organized in terms of
infinite-dimensional representations of $su(2)$. We argue that this gives
rise to a free-field realization of the $\su(2)_{-1/2}$ model.

In the subsequent section, we confirm this identification
by comparing with the correlators of the $\su(2)_{-1/2}$ model determined
using the Knizhnik-Zamolodchikov (KZ) equation.

In the conceptually most important Section 6, we study the
fusion rules in the $\beta\gamma$ system, and hence,
in the $\su(2)_{-1/2}$ model. We show that fusions of the basic
twists generate fields of arbitrarily large negative dimensions.
This can be interpreted
in terms of the deeper twists in the $\beta\gamma$ system, or in
terms of the action of the spectral
flow in the $\su(2)_{-1/2}$ WZW model. This is in sharp contrast to
most expectations in the literature, where the existence of the modular
invariant in \cite{KK,Lu}
has been interpreted (at least implicitly) as a ``proof'' that the
$\su(2)_{-1/2}$ model is a  rational CFT. These observations have some
overlap with the results obtained recently by Maldacena and Ooguri
\cite{MO} on the $\widehat{sl}(2,\RR)$ WZW model, 
and by Gaberdiel on the $\su(2)_{-4/3}$ WZW model \cite{Gab}.
In particular, the addition of extra fields under the spectral flow
is reminiscent of the solution to the $\widehat{sl}(2,\RR)$ WZW model
\cite{MO}. However, the two models are different, and the appearance of
continuous representations in \cite{MO} is not replicated here.
Finally, we show that the modular invariant partition
function written in \cite{KK} {\em is} compatible with the infinite
operator content that we encounter, 
provided the characters are properly interpreted.

Section 7 contains some concluding remarks.
The issue of logarithms alluded to above is
the subject of our forthcoming paper \cite{LMRS}.

\section{The $\bg$ system}

\subsection{Generalities}

The bosonic $\beta\gamma$ system or $c=-1$ ghost theory is defined by the
(first order) action (\ref{betaaction}). Here we focus on the left-moving 
sector and omit the subscript $L$ for simplicity.
The functional integral is clearly not
well-defined. How formal manipulations of the model
can give rise to a meaningful physical theory,
is to a large extent the essence of this paper.
At first sight, a formal treatment of the $\beta\gamma$ system produces
very simple results. The elementary correlators, obtained
through analytic continuation of the Gaussian integrals, read
\be
    \L\beta(z)\gamma(w)\R= -\L\gamma(z)\beta(w) \R= {1\over z-w}
\label{bg}
\ee
The stress-energy tensor is
\be
T= {1\over 2}\left(:\beta\d\gamma:-:\d \beta \gamma:\right)
\label{T}
\ee
and leads to the central charge $c=-1$, while $\beta$ and $\gamma$ both
have weight $h={1\over 2}$. There is also an obvious $u(1)$ charge
with current
\ben
J^3=-{1\over 2}:\gamma\beta:
\label{J3}
\een
with respect to which $\beta$ has charge $1/2$ and $\gamma$ charge $-1/2$.

The ghost fields can be periodic (NS sector, $p=0$)
or anti-periodic (R sector, $p=1$), and have the mode expansions:
\be
   \beta(z)= \sum_{n\in\ZZ} z^{-n-1-p/2}\beta_{n+(1+p)/2}\qquad
    \gamma(z)= \sum_{n\in\ZZ} z^{-n-1+p/2}\gamma_{n+(1-p)/2}
\ee
The associated commutators read
\be
   \left[\beta_s,\gamma_t\right]=\delta_{s+t,0}
\ee
In the NS sector ($p=0$) the ground state is defined by
$\beta_{r}|\varphi_0\R =\gamma_{r}|\varphi_0\R=0$ for $r>0$ and
$r\in\NN+1/2$. The normal-ordered Hamiltonian is then
\be
L_0=\sum_{n\geq 0}
(n+1/2)\left(\gamma_{-n-1/2}\beta_{n+1/2}-\beta_{-n-1/2}
\gamma_{n+1/2}\right)
\ee
In the R sector ($p=1$), the Hamiltonian reads
\be
L_0=-{1\over 8}+\sum_{n>0}n\left(\gamma_{-n}\beta_n-\beta_{-n}\gamma_n\right)
\ee
Because of the existence of zero  modes, $\beta_0$ and $\gamma_0$,
different choices can be made for the ground state. These choices have
deep implications in terms of the $su(2)$ content of the theory.
This will be addressed in the context of the spectral flow in Section 6.
One of the simplest choices is to demand that
\be
     \beta_{n+1}|\varphi_1\R = \gamma_{n}|\varphi_1\R = 0,\ \ \ \ n\geq 0
\ee
in which case $\beta_0|\varphi_1\R\not=0$.
The vacuum state is infinitely degenerate since
$\beta_0^N|\varphi_1\R$ is a vacuum state for any $N$. The
states are naturally organized in terms of the parity of the number
of acting $\beta$ modes. This infinite number of states will thus
split into two sequences of definite parity, and each sequence will
eventually be associated to an infinite $su(2)$ representation.

The $u(1)$ charge in the R sector reads
\be
   J^3_0=-{1\over 4}\left(\beta_0\gamma_0+\gamma_0\beta_0\right)
    ={1\over 4}-{1\over 2}\beta_0\gamma_0
\ee
The $u(1)$ charge of the state
$|\varphi_1\R$ is $J_0^3={1\over 4}$, while the $u(1)$ charge of
$\beta_0^N|\varphi_1\R$ is $J_0^3={1\over 4}+{N\over 2}$.
One could as well exchange the roles of $\beta$ and
$\gamma$, and define the ground state in the R sector through
\be \label{llwcond}
   \gamma_{n+1}|\varphi_1\R = \beta_{n}|\varphi_1\R = 0,\ \ \ \ n\geq 0
\ee
In this case, $|\varphi_1\R$  has charge $J_0^3=-{1\over 4}$, and the
states $ \gamma_0^N|\varphi_1\R$ charge  $J_0^3=-{1\over 4}-{N\over 2}$.
Finally, it is also possible that the ground state is
annihilated neither by $\beta_0$ nor $\gamma_0$. This would lead to a tower of
states with values of $J_0^3$ extending infinitely in both directions.

In the rest of this section, we choose the ground state of the NS and R
sector to obey the highest-weight conditions
\be
   \beta_{n+(1+p)/2}|\varphi_p\R = \gamma_{n+(1-p)/2}|\varphi_p\R = 0,
    \ \ \ \ n\geq 0
\label{hwcond}
\ee
The character, $\chi$, of the Verma module is obtained by counting all
possible applications of the lowering operators $\beta_{n+(1+p)/2}$
and $\gamma_{n+(1-p)/2}$ with $n<0$. Keeping track of the
powers of $\beta$ with a factor $y$ and those of $\gamma$ with a
factor $y^{-1}$, we thus have
\be \label{cartr}
\chi_p (q,y) = q^{1/24} q^{-\frac{1}{8}(1-(1-p)^2)}
\prod_{n=0}^\infty \frac{1}{(1+yq^{n+\frac{1}{2}(1-p)}) }
\frac{1}{(1+y^{-1} q^{n+\frac{1}{2}(1+p)})}
\ee
In the expansion of the infinite product, the coefficient of $q^0$
(which can occur only for
$p=1$) is $\sum_{m\geq 0} y^m$. This reflects the infinite degeneracy
of the vacuum. 

Up to this point, the $c=-1$ theory defined with the above highest-weight
conditions resembles the
$c=2$ theory of a complex boson \cite{LudwigNPB519}. This is
not surprising. For instance, the formal partition function of the
$\beta\gamma$ system in the RR sector on the torus would read
$Z\propto 1/\hbox{det }\Delta$, where $\Delta$ is the Laplacian. 
This extends to 
the partition function when NS and R sectors are combined. Recall that 
the determinant of the Laplacian acting
on functions with boundary conditions (the complex variable on the torus is
denoted $\zeta$ to avoid confusion)

\ben
   f(\zeta+\omega_{1})=-e^{2i\pi\alpha}f(\zeta),\ \ \ \
   f(\zeta+\omega_{2})=-e^{2i\pi\beta}f(\zeta),\ \ \ \
   \tau=\omega_{2}/\omega_{1}
\een
   takes the form (cf. the character (\ref{cartr}) with $\alpha=p/2$ and
$e^{2\pi i \beta}=y$):
\be
    {1\over \hbox{det}\Delta_{\alpha+1/2,\beta+1/2}}
    \ =\ q^{1/24}q^{-\alpha^{2}/2}
    \prod_{n=0}^{\infty}
     {1\over (1+e^{2i\pi\beta}q^{n+1/2-\alpha})}
     {1\over(1+e^{-2i\pi\beta}q^{n+1/2+\alpha})}\times(c.c.)
\ee
The associated  partition function reads ($\beta$ is denoted $z$ for 
future convenience)
\be
   Z=\left({1\over \hbox{det
    }\Delta_{1/2,z+1/2}}+{1\over \hbox{det
    }\Delta_{1/2,z}} + {1\over \hbox{det
    }\Delta_{0,z+1/2}} + {1\over \hbox{det }\Delta_{0,z}}\right)\label{Zu}
\ee
The ``natural case'' is $z=0$, and corresponds to
what one would like the partition function to be,
based on formal functional integral calculations with the $\beta\gamma$
action.
This coincides, of course, with the partition function of a
non-compactified complex boson. The infinite degeneracy of the vacuum in
the operator formulation can then be interpreted as a manifestation of
the zero mode of the Laplacian.
Partition functions (characters) in different sectors can be matched,
and it is tempting to argue that the $\beta\gamma$ system and the
non-compactified complex boson are essentially equivalent,
the former being a sort of twisted version of the latter  \cite{LudwigNPB519}.

We shall see in the following that this is not true. In
particular, the $\beta\gamma$ system defined through the highest-weight
conditions (\ref{hwcond}) is a complicated CFT where the divergences present
in the functional integral give rise to a spectrum unbounded from below,
and the partition function (\ref{Zu}) will not be the final answer.
The formal similarity with the non-compactified complex boson is indeed
only formal and hides a subtle resummation of this spectrum.

To uncover the physical content of the $\beta\gamma$ system requires
a deeper understanding of the fields in the R sector, to which we now turn.

\subsection{Twist fields and monodromy considerations}

The R sector corresponds to anti-periodic boundary conditions for
the $\beta$ and $\gamma$ fields on the plane. These boundary
conditions are generated by twist fields, for which we demand
the local monodromy conditions
\be
   \beta(z)\tau(1)\propto (z-z_{1})^{-1/2},\ \ \ \ \gamma(z)\tau(1)\propto
    (z-z_{1})^{-1/2}
\label{mon}
\ee
The square root singularities multiply ``excited'' twist fields.
Here and below we use the abbreviation $\tau(i)=\tau(z_i)$.

Let us calculate the conformal weight of the twist
fields. For this, we consider the ratio of correlators
\be
    g_2(z,w)={\langle \beta(z)\gamma(w)\tau(2)\tau(1)\rangle
\over \langle \tau(2)\tau(1)\rangle}
\ee
in the limit where $z_{1}=0$ and $z_{2}=\infty$. It follows that
\be
    g_2(z,w)=z^{-1/2}w^{-1/2} \,{Az+(1-A)w\over z-w}
\ee
where the square root part is due to (\ref{mon}), while
the meromorphic function follows from
the constraint that $g_2(z,w)\propto {(z-w)}^{-1}$ as $z\rightarrow w$,
cf. (\ref{bg}). $A$ is unconstrained and is left as a
free parameter. Its meaning is immediately elucidated when considering
the $u(1)$ current (\ref{J3}). Namely, from $g_2(z,w)$ we find the OPE
\be
   {\langle :\beta\gamma:(w)\tau(2)\tau(1)\rangle
    \over \langle \tau(2)\tau(1)\rangle}={A-1/2\over w}+\ldots
\ee
as $w\rightarrow z_{1}$, and thus, the $u(1)$ charge of the twist field
at $z_{1}$ is $J_{0}^{3}=-{1\over 2}(A-1/2)$. It must be opposite to the
$u(1)$ charge of the conjugate twist field at $z_{2}$ in order for the
correlator not to vanish.  Thus, despite the notation
used above, the two twist fields within the correlator may be different
-- the twist field is not unique.

The $u(1)$ charge does not affect the conformal weight, in contrast
to what would happen in the fermionic $\eta\xi$ system, for example.
To see this, we use the stress-energy tensor (\ref{T}). Evaluating the leading
singularity as $w\rightarrow z_{1}$ along the same lines as above, one finds
\be
{\langle :\partial\beta\gamma:(w)\tau(2)\tau(1)\rangle
   \over \langle \tau(2)\tau(1)\rangle}=\left({3\over 8}-{A\over
   2}\right){1\over w^{2}}+\ldots
\ee
and
\be
{\langle :\partial\gamma\beta:(w)\tau(2)\tau(1)\rangle
\over \langle \tau(2)\tau(1)\rangle}=-\left({3\over 8}-{(1-A)\over
2}\right){1\over w^{2}}+\ldots
\ee
It follows that $h=-{1\over 8}$, independently of the value of $A$.

In theories of free
Majorana and Dirac fermions, or of free bosons, monodromy conditions like
(\ref{mon}) define the twist fields uniquely. That is not so in the
$\beta\gamma$ system: although the conformal weight of $\tau$ is
fixed uniquely by the local monodromy ($h=-{1\over 8}$), an ambiguity
related to the $u(1)$ charge remains.

\subsection{Free-field representation: $\bg$ twists vs $\ex$ twists}

To understand better the role of the $u(1)$ charge, it is convenient
to introduce the free-field representation
\be
    \beta=e^{-i\phi}\eta,\ \ \ \ \gamma=e^{i\phi} \partial\xi
\ee
where $\phi$ is a free boson with negative metric:
\be
    \langle\phi(z)\phi(w)\rangle=+\ln(z-w)
\ee
We use an implicit notation where vertex operators like 
$e^{i\phi}$ are normal ordered.
$\eta$ and $\xi$ are fermions of weight $h=1$ and $h=0$, respectively, obeying
\be
    \langle\eta(z)\xi(w)\rangle=\langle\xi(z)\eta(w)\rangle={1\over z-w}
\ee
Since the exponentials in the free-field representation have weight
$h=-{1\over 2}$, $\beta$ and $\gamma$ have weight $h={1\over 2}$, as required.
Recall that the $\eta\xi$ system has stress-energy tensor
$T=:\partial\xi\eta:$, and central charge $c=-2$.

It is important to notice that the $\beta\gamma$ system has a single
$u(1)$ charge, while this representation allows for two of them. The
free-field representation of $J^3$ involves the $u(1)$ charge
of the boson only, $J^3={1\over 2}i\partial\phi$, but
there is also the $u(1)$ charge of the $\eta\xi$ system, $j=:\xi\eta:$.

The free-field representation makes it clear that to  create a branch
for $\beta$ and $\gamma$, we have a {\em continuum} of possibilities.
Indeed, introduce a twist\footnote{The difference in the two OPEs
stems form the fact that $\xi$ and $\eta$ have different
dimensions. Twists are more easily defined in the symplectic fermion
theory, using two fermionic currents.}
for the fermionic fields:
\be
   \partial \xi(z)\st_\lambda(w)\propto (z-w)^{\lambda-1},\ \ \ \
     \eta(z)\st_\lambda(w)\propto (z-w)^{-\lambda}\label{iniope}
\ee
or more precisely
\be
   \partial\xi(z e^{2\pi i})\,  \st_\lambda(w)= e^{2\pi
     i(\lambda-1)}\, \partial\xi(z)\, \st_\lambda(w) ,\qquad
     \eta(z e^{2\pi i})\,  \st_\lambda(w)= e^{-2\pi i\lambda}\, \eta(z)\,
     \st_\lambda(w)
\ee
Such twists have been studied before \cite{Saleur,Kausch}.
Their conformal dimension and $\eta\xi$ charge (i.e., $j_0$ eigenvalue)
are given by
\be
    h_{\st_\lambda}= -{\la(1-\la)\over 2}\; ,\qquad q_{\st_\lambda}= \la
\ee
In particular, $\xi$ and $\eta$ have $\eta\xi$ charges $+1$ and $-1$,
respectively. To complement this, we can introduce a ``magnetic'' charge
operator, $e^{i\alpha(\phi-\bar{\phi})}$, in the free-boson theory, and
select
$\alpha$ in order for $\beta$ and $\gamma$ to be  anti-periodic when acting
on $\st_\la e^{i\alpha(\phi-\bar{\phi})}$. For instance,
\be
    \beta(ze^{2\pi i})\, e^{i\alpha \phi(w)}\, \st_\lambda(w)= e^{2i\pi
    (\alpha-\lambda)\, }\beta(z)\, e^{i\alpha
    \phi(w)}\st_\lambda(w)
\ee
and we want the phase to be $e^{-i \pi  }$, so that
\be
    2\alpha-2\lambda=-1\label{allarel}
\ee
One then has
\be
    \gamma(ze^{2\pi i})\, e^{i\alpha \phi(w)}\, \st_\lambda(w)= e^{2i\pi
    (-\alpha+\lambda-1)\, }\gamma(z)\, e^{i\alpha
    \phi(w)}\st_\lambda(w)
\ee
with a corresponding  phase  $e^{-i \pi  }$ as well.
In the following, $\alpha$ and $\lambda$ are
always assumed to be related by (\ref{allarel}).
The expression for the full twist field follows:
\be
{\tilde\tau}_{\lambda} =\st_{\lambda}
e^{i(\lambda-{1\over 2})(\phi-\bar{\phi})}
\ee
Its dimension is
\be
h={-\lambda(1-\lambda)\over 2}-{(1-2\lambda)^2\over 8}=-{1\over 8}
\ee
independent of $\lambda$, as desired. As already mentioned, in the free-field
representation the $u(1)$ current of the $\beta\gamma$ system reads
$J^{3}={1\over 2}i\partial\phi$. Thus, the charge of the
twist field is $J^3_0=-{1\over 2}\left(\lambda-{1\over 2}\right)$, and
$\lambda$ coincides with the parameter $A$ introduced above.
Conjugate twist fields are obtained by replacing $\lambda$ by $1-\lambda$.

Let us now try to identify particular $c=-2$ twist fields.
Given its dimension and charge, we can set $\st_0=I$. Note that
$\st_1$ also has dimension 0 but it has $\eta\xi$ charge
$+1$, hence it cannot be equivalent to the identity field. But since
it has the same dimension and charge as $\xi$, it is natural to set
$\st_1=\xi$. There are two other values of $\la$ that
suggest simple identifications: $\st_{-1}$ and $\st_2$ both have
dimension 1, and their charges are $-1$ and $+2$, respectively.
It is thus natural to set $\st_{-1}=\eta$ and $\st_2=\xi\d\xi$.
These identifications are supported by the following, slightly
different OPEs, compatible with charge conservation:
\be
    \eta(z)\st_\la(w)\sim  (z-w)^{-\la} \;\st_{\la-1}(w),\qquad
    \d\xi(z)\st_\la(w)\sim
     { (z-w)^{\la-1}} \;\st_{\la+1}(w)\label{twistope}
\ee
These OPEs in turn (as well as the charge and dimension
assignments) are compatible with the
following more general identifications ($n>0$):
\begin{eqnarray}
    \st_n =\xi \d\xi \cdots \d^{n-1}\xi \, \qquad &{\rm with }&\quad
     h_{\st_n}={n(n-1)\over 2},\qquad
     q_{\st_n}=n\nonumber\\
     \st_{-n} =\eta \d\eta \cdots \d^{n-1}\eta \, \qquad &{\rm with}&\quad
     h_{\st_{-n}}={n(n+1)\over 2},\qquad q_{\st_{-n}}=-n
\label{identification}
\end{eqnarray}
It is now crucial to recall that there are different choices of $c=-2$
theories, which will lead to different realizations of ``the''
$\beta\gamma$ system. Here, we will choose to use not the $\eta\xi$
system but rather the $\eta\d\xi$ system\footnote{This choice is justified
by the observation that the $\widehat{su}(2)_{-1/2}$ symmetry
constructed in the $\beta\gamma$ system
commutes with a BRST charge $Q_\eta=\oint \eta$ in the $\eta\xi\phi$
free-field representation. How logarithms may appear when going beyond this
choice, will be the subject of our forthcoming paper \cite{LMRS}.}
(though, for simplicity refer to it as the $\eta\xi$ system).
In the terminology of \cite{Friedan,Kausch}, we thus work in the
{\it small algebra}, generated by the modes of $\eta$ and $\d\xi$.
We denote the corresponding twist fields by
$\s_\la$. Instead of (\ref{identification}), for instance, we then have
\bea
   \s_n =\d\xi \cdots \d^{n-1}\xi \, \qquad &{\rm with }&\quad
   h_{\s_n}={n(n-1)\over 2},\qquad q_{\s_n}=n-1\nn
   \s_{-n} =\eta \d\eta \cdots \d^{n-1}\eta \, \qquad &{\rm with}&
    \quad h_{\s_{-n}}={n(n+1)\over 2},\qquad q_{\s_{-n}}=-n
\label{identificationi}
\eea
Particular examples are $\s_1=I$ and $\s_2=\d\xi$. These identifications
are compatible with the OPEs (\ref{iniope}) for $n\neq 0,1$. For
$n=0,1$, since $\sigma_0=\sigma_1=I$, the OPEs have to be changed
by the addition of integer powers of $z-w$.

The definition of the twist fields in the $\beta\gamma$ system is now
\be
   \tau_{\lambda}\equiv \s_{\lambda}
    e^{i(\lambda-{1\over 2})(\phi-\bar{\phi})}
\label{iden}
\ee
($\bar{\phi}$ is omitted in the following).
With this definition, the OPEs of twists in the $\beta\gamma$ system
are generically satisfied, except for some special values of $\lambda$.
For instance, one has $\beta(z)\tau_0(w)\sim (z-w)^{-1/2}$ and
$\gamma(z)\tau_0(w)\sim (z-w)^{1/2}$. Similarly, we have
$\beta(z)\tau_1(w)\sim (z-w)^{1/2}$ and $\gamma(z)\tau_1(w)\sim
(z-w)^{-1/2}$.  Otherwise $\beta(z)\tau_{\pm n} (w)\sim (z-w)^{-1/2}$ and
$\gamma(z)\tau_{\pm n}(w)\sim (z-w)^{-1/2}$.
The marginal cases  can be obtained within
our formalism as limits of the generic cases. For instance, the
four-point function $\langle \beta\gamma\tau_1\tau_0\rangle/
\langle\tau_1\tau_0\rangle$ goes as
$(z/w)^{1/2}(z-w)^{-1}$, and corresponds to the particular case $A=1$.

\subsection{Twist correlators}

We now turn to the calculation of the four-point functions of twist fields.
There are several ways to tackle this question. A possibility is to
follow the lines of \cite{Dixon,Saleur}, that is, to implement local as
well as global monodromy constraints in the $\beta\gamma$ system directly.
The starting point would be the ratio of correlators
\be
   g_4(u,w)={\langle \beta(u)\gamma(w)\tau(4)\tau(3)\tau(2)\tau(1)\rangle
    \over \langle \tau(4)\tau(3)\tau(2)\tau(1)\rangle}
\ee
(the charge labels being introduced below).
Using monodromy arguments based on (\ref{mon}) and the result (\ref{bg}),
would, however, not lead to a complete determination of $g_4$.
There would, once again, remain ambiguities because of the
$u(1)$ charge of the twist fields. If we were to assume, for
instance, that $\tau(1)$ and $\tau(3)$ on the one hand, and
$\tau(2)$ and $\tau(4)$ on the other, had the same charge,
it would follow that
\begin{eqnarray}
     &&g_4(u,w)\ =\ \prod_{i}(u-z_{i})^{-1/2}(w-z_{i})^{-1/2}\nn
     &\times&\left[{A(u-z_{1})(u-z_{3})(w-z_{2})(w-z_{4})\over (u-w)}
       +{B(u-z_{2})(u-z_{4})(w-z_{1})(w-z_{3})\over (u-w)}
  +C(z_i,\bar{z}_i)\right]
\label{begamono}
\end{eqnarray}
with the only constraint $A+B=1$ ($A,B$ are constants, while $C$ is 
a constant with respect to $z$ and $w$ but 
depends on the other arguments). The choice of $A$ (and hence of
$B$) would correspond to a choice of $u(1)$ charge for
each of the twist fields, exactly as in the case of the two-point function.

In fact, to evaluate this correlator, it is easier to use the free-field
representation.  The bosonic part of the twist correlator is then
evaluated straightforwardly  using Wick's theorem. As for the fermionic
part, we can use the method  of \cite{Dixon,Saleur}. Let us indicate how.
The starting point is
\be
   g_4(u,w)={\langle \eta(u)\partial\xi(w)
  \sigma_{1+\rho}(4)\sigma_{\nu}(3)\sigma_{1+\mu}(2)\sigma_{\lambda}(1)\rangle
    \over \langle \sigma_{1+\rho}(4)\sigma_{\nu}(3)
   \sigma_{1+\mu}(2)\sigma_{\lambda}(1)\rangle}
\ee
with $\lambda+\mu+\nu+\rho=0$. Local monodromy constraints lead to
the introduction of the two forms
\begin{eqnarray}
        \omega_{1}(u)&=&(u-z_{1})^{-\lambda}(u-z_{2})^{-1-\mu}(u-z_{3})^{-\nu}
        (u-z_{4})^{-1-\rho}\nonumber\\
     \omega_{2}(w)&=&(w-z_{1})^{\lambda-1}(w-z_{2})^{\mu}
    (w-z_{3})^{\nu-1}(w-z_{4})^{\rho}
\end{eqnarray}
$g_4(u,w)$ can then be written as the product $\omega_{1}\omega_{2}$ times
an analytic function of
$u,w,z_{i}$, which can be determined uniquely.
Sending $z\rightarrow w$, extracting the stress-energy tensor from the
$\eta\partial\xi$ OPE, and using the OPE of $T$ with primary fields,
leads (after setting  $z_{1}=0$, $z_{2}=z$, $z_{3}=1$ and $z_{4}=\infty$)
to a differential equation in $z$ for the object
\be
        G(z,\bar{z})=\hbox{lim}_{z_{\infty}\rightarrow\infty}
      |z_{\infty}|^{\mu^{2}+\nu^{2}+\lambda\nu+\lambda\mu+\nu\mu-\nu}
     \langle\sigma_{1+\rho}(z_{\infty},\bar{z}_{\infty})
    \sigma_{\nu}(1,1)\sigma_{1+\mu}(z,\bar{z})\sigma_{\lambda}(0,0)\rangle
\ee
In the related free-boson problem, it is referred to as the quantum
correlator.
The second part of the calculation involves the function
\be
   h_4(u,\bar{w})={\langle
    \eta(u)\bar{\partial}\bar{\xi}(\bar{w})
  \sigma_{1+\rho}(4)    \sigma_{\nu}(3)
  \sigma_{1+\mu}(2)\sigma_{\lambda}(1)\rangle
    \over \langle \sigma_{1+\rho}(4)
    \sigma_{\nu}(3)\sigma_{1+\mu}(2)\sigma_{\lambda}(1)\rangle}
\ee
While we had the OPE $\xi(u)\sigma_{\lambda}(w,\bar{w})\propto
(u-w)^{\lambda}$, we now need
$\bar{\xi}(\bar{u})\sigma_{\lambda}(w,\bar{w})\propto
(\bar{u}-\bar{w})^{-\lambda+1}$. It follows that
$h_4(u,\bar{w})=\omega_{1}(u)\bar{\omega}_{1}(\bar{w})D(z_{i},\bar{z}_{i})$,
where $D$ is a constant (with respect to $z$ and $w$).
The final step is to demand that $\xi$ is uniquely defined
when going around contours that encircle the four twist fields. This involves
the integral of $\omega_{1}$ along various cycles.
For the benefit of our presentation, we focus on the ones that are
related to the fundamental integral
\be
        \int_{0}^{1} dy y^{-\lambda}(1-y)^{-1-\mu}(1-zy)^{-\nu}
        ={\Gamma(1-\lambda)\Gamma(-\mu)\over\Gamma(1-\lambda-\mu)}
        F(\nu,1-\lambda;1-\lambda-\mu;z)
\label{2F1}
\ee
This integral formula is valid for Re$(1-\lambda)$,\ Re$(-\mu)>0$ and
$|z|<1$. Here
$F$ denotes the hypergeometric function ${}_2F_1$. The final result is
\be
    G\ \propto\ |z|^{\mu(\lambda-1)}|1-z|^{\nu(1+\mu)}\left[
    F(\nu,1-\lambda;1-\lambda-\mu;z) F(\nu,1-\lambda;1-\lambda-\mu;1-\bar{z})+
        (c.c.)\right]
\ee
Going back to the physical correlator, one has
\begin{eqnarray}
    &&\langle\sigma_{1+\rho}(4)
     \sigma_{\nu}(3)\sigma_{1+\mu}(2)\sigma_{\lambda}(1)\rangle\nonumber\\
    &\propto&|z_{12}|^{-h_1-h_2+h_3+h_4}
     |z_{13}|^{-2h_3}|z_{14}|^{-h_1+h_2+h_3-h_4}|z_{24}|^{h_1-h_2-h_3-h_4}
     \nonumber\\
    &\times&|z|^{\nu(1-\lambda-\mu-\nu)}|1-z|^{\nu(1+\mu)}
     \left[F(\nu,1-\lambda;1-\lambda-\mu;z)
    F(\nu,1-\lambda;1-\lambda-\mu;1-\bar{z})+(c.c.)\right]
\label{fourpt}
\end{eqnarray}
with $z_{ij}=z_i-z_j$.
Note again that we are interested in values of
$\lambda,\mu,\nu,\rho$ for which the OPEs initially written may not
hold exactly, and differ from the ones used in this derivation
by integer powers of $z-w$. Our practical definition of the
twist fields will be through the four-point function, which we will
demand to always be given by (\ref{fourpt}). If it so happens that the
hypergeometric function is then ill-defined, the pole
may formally be factored out, and a new hypergeometric function,
this time well-defined, is substituted. This simple procedure
corresponds to supplementing (\ref{2F1}) with an integral formula
valid for other values of the parameters $\lambda,\mu,\nu,\rho$,
and implementing it in the expression for the four-point
function (\ref{fourpt}). See also the expansion discussed in Section 5.1.

The above correlator is compatible with all the identifications proposed
previously, in particular with setting
$\sigma_{n}=\,:\partial\xi\ldots\partial^{n-1}\xi:$
and $\sigma_{-n}=\,:\eta\partial\eta\ldots\partial^{n-1}\eta:$.
To illustrate this, we consider
\be
    \s_{3}=:\partial\xi\partial^2\xi:,~~~\ \ \ \s_{-2}=
     :\eta\partial\eta:
\ee
and compute the four-point function
\be
    G=\left\langle:\partial\xi\partial^{2}\xi:(4)
       :\partial\xi\partial^{2}\xi:(3):\eta\partial\eta:(2)
          :\eta\partial\eta:(1)\times (c.c.)\right\rangle
\ee
We should then check that it coincides with the corresponding case of our
general result:
\be
    G\ \propto\ |z_{13}z_{24}|^{-12}|z(1-z)|^{-12}
     \left[F(3,-2;1;z)F(3,-2;1;1-\bar{z})+(c.c.)\right]
\ee
Since $F(3,-2;1;z)=1-6z+6z^{2}$, we end up with
\be
    G\ \propto\
     |z_{13}z_{24}|^{-12}|z(1-z)|^{-12}(1-6z+6z^2)(1-6\bar{z}+6\bar{z}^{2})
\ee

Finally, we can go back to the free-field representation of the
$\beta\gamma$ system and obtain
\begin{eqnarray}
    \langle\tau_{1+\rho}(4)
     \tau_{\nu}(3)\tau_{1+\mu}(2)\tau_{\lambda}(1)\rangle\propto
     |z_{13}z_{24}|^{1/2} |z(1-z)|^{1/2}\nonumber\\
    |z|^{-\lambda-\mu}|1-z|^{-\lambda-\rho}
    \left[F(1-\lambda,\lambda;1;z)F(1-\lambda,\lambda;1;1-\bar{z})
    +(c.c.)\right]\label{twistcorr}
\end{eqnarray}
A particularly simple case is
\begin{eqnarray}
    &&\langle\tau_{1-\lambda}(4)
     \tau_{\lambda}(3)\tau_{1-\lambda}(2)\tau_{\lambda}(1)\rangle\nonumber\\
    &\propto&|z_{13}z_{24}|^{1/2} |z(1-z)|^{1/2}
    \left[F(1-\lambda,\lambda;1;z)
    F(1-\lambda,\lambda;1;1-\bar{z})+(c.c.)\right]
\end{eqnarray}
This computation demonstrates the equivalence of the twist-field
correlators based on the monodromy definition, and the explicit
representation in terms of fermions obtained in the previous subsection.

\section{The $\su(2)_{-1/2}$ WZW model}

In this section we review some aspects of the $\su(2)_{-1/2}$
WZW model. Its relation to the $\bg$ system is discussed in the
subsequent sections.

\subsection{Admissible representations of the $\su(2)_{-1/2}$ model}

A fractional level $k$, for the $\su(2)_{k}$ algebra, is said to be
admissible if $k=t/u$, with $t\in\ZZ$ and
$u\in \NN$ relative prime, and $t+2u-2\geq 0$ \cite{KK} (see also
\cite{DMS}, Section 18.6). The admissible
$\su(2)_k$ representations are then  characterized by those spins $j$
which can be parameterized by two non-negative integers $r$ and $s$ as
\be
\label{rsnot}
    2 j+1= r-(k+2)s,\ \qquad 1\leq r\leq t+2u-1,\ \qquad 0\leq s\leq u-1
\ee
In the present case $t+2u-1=u=2$, so that $r=1,2$ and $s=0,1$. Thus, there
are four admissible representations, with
$(r,s)= (1,0),\; (2,0),\; (1,1),\; (2,1)$.
With this ordering, they correspond to
\be
\label{spi}
    j=0\;,\quad \frac{1}{2}\;,\quad -\frac{3}{4}\;,\quad -\frac{1}{4}
\ee
Their highest-weight states have conformal dimension $h_j$ given by
\be
    h_j = \frac{j(j+1)}{k+2}
\ee
The admissible conformal dimensions are thus
\be \label{didi}
    h_0=0\;, \quad h_{1/2}=\frac{1}{2}\;,\quad h_{-3/4}=h_{-1/4}=-
     \frac{1}{8}
\ee
The characters of the admissible representations are given by
\be
  \chi_{j}^\epsilon (q,y)= {\rm Tr}_{D_j^\epsilon} \,q^{L_0-c/24}\, y^{2J^3_0}
\label{trch}
\ee
$D_j^\epsilon$ denotes the representation of spin $j$ with $\epsilon = \pm$ 
for highest or lowest weight, while
\be
   q=e^{2\pi i \tau}, \; \qquad y=e^{2\pi i z}
\ee
This trace can be summed and for
the highest-weight representations of Kac and Wakimoto, we get \cite{KK} 
(see e.g. \cite{DMS}, Eq. 18.185)
\be
\label{cara}
   \chi_{j}^+(\tau,z) =
    \frac{\Theta_{b_+}^{(d)}(z/u;\tau)-\Theta_{b_-}^{(d)}(z/u;\tau)}{
    \Theta_{1}^{(2)}(z;\tau)-\Theta_{-1}^{(2)}(z;\tau)}
\ee
Here we have used the notation
\be
    \Theta_b^{(d)} = \sum_{l\in\ZZ+\frac{b}{2d}} q^{dl^2} y^{-dl}
\ee
with $k=-1/2 = t/u$, $u=2$, $d=u^2(k+2)=6$, and
\be
    b_{\pm} = u[\pm(r)-(k+2) s]
\ee
For the identity, for example, we have
$b_\pm = \pm 2$, while for the field with $j=-1/4$, we have $b_+=1$
and $b_-=-7$.

The $\su(2)_{-1/2}$ admissible characters turn out to have simple
expressions in terms of the usual theta functions:
\begin{eqnarray}
        \chi_{0}(\tau,z)&=&{1\over 2}\left[
{\eta\over\theta_{4}(\tau,z)}+{\eta\over\theta_{3}(\tau,z)}\right]\nonumber\\
         \chi_{1/2}(\tau,z)&=&{1\over 2}\left[
{\eta\over\theta_{4}(\tau,z)}-{\eta\over\theta_{3}(\tau,z)}\right]\nonumber\\
         \chi_{-1/4}^{+}(\tau,z)&=&{1\over 2}\left[ {\eta\over
i\theta_{1}(\tau,z)}+{\eta\over  \theta_{2}(\tau,z)}\right]\nonumber\\
         \chi_{-3/4}^{+}(\tau,z)&=&{1\over 2}\left[ {\eta\over
i\theta_{1}(\tau,z)}-{\eta\over \theta_{2}(\tau,z)}\right]
\label{adtheta}
        \end{eqnarray}
These admissible character functions close under the modular
group \cite{KK}. For this reason, the $\su(2)_{-1/2}$ WZW model is
often said to be a rational CFT (and likewise for more general
admissible level
$\su(2)_{k}$ theories). This statement will be re-evaluated in Section 6
where we are more careful about the convergence regions of the
traces (\ref{trch}) when summed to the character functions (\ref{cara})
and (\ref{adtheta}).

The associated diagonal modular invariant\footnote{This is not the sole
invariant we can write, however. Given that charge conjugation
(defined from the square of the modular $S$ matrix) relates $j$ to
$-j-1$, we also have the charge conjugate version of the diagonal invariant:
$$
   Z^{\rm C}=|\chi_{0}|^2+|\chi_{1/2}|^2+\chi_{-1/4}^+\chi_{-3/4}^{+*}
    +\chi_{-3/4}^+\chi_{-1/4}^{+*}
$$}
is given by\footnote{If the variable $z$ transforms as
$z\tau /(a\tau+b)$ as $\tau$ is changed to $(a\tau+b)/(c\tau+d)$,
there is no prefactor in this partition function. If, on the other
hand, $z$ is changed to $z/(c\tau+d)$, there is the prefactor
$\exp[2\pi {\rm Im}(z)^{2}/{\rm Im}(\tau)]$, cf. appendix B of \cite{MO}.}
\be
   Z(\tau,z)\ =\ |\chi_{0}(\tau,z)|^2+|\chi_{-1/4}^+(\tau,z)|^2
   +|\chi_{1/2}(\tau,z)|^2+|\chi_{-3/4}^+(\tau,z)|^2\ =\ \sum_{i=1}^{4}
    {|\eta(\tau)|^{2}\over |\theta_{i}(\tau,z)|^{2}}
\label{Ze}
\ee
This can be shown to coincice with (\ref{Zu}). 

We stress that in the present work, when we refer to highest- or lowest-weight
representations, the qualitative ``highest'' or ``lowest'' refers to the
grade-zero $su(2)$ representation: a highest-weight state is annihilated
by $J^+_0$ while a lowest-weight state is annihilated by $J^-_0$.
The superscript, $+$ or $-$, on the character indicates that the
representation is a highest-weight or lowest-weight representation,
respectively. Note that for $j=0,1/2$, the representation is both
a highest- and a lowest-weight representation. On the other hand,
both types of representations are affine
highest-weight representations, that is, they are annihilated
by the action of $J^+_n\, , J^-_n$ and $J^3_n$ for $n>0$.
The functional form above is not a property which relies upon
choosing the
representations to be $su(2)$ highest- or lowest-weight representations.
For example, $\chi_{-3/4}^{+}= -\chi_{-1/4}^{-}$ and
$\chi_{-1/4}^{+}= -\chi_{-3/4}^{-}$ (see below).
In other words, the admissible representations could be regarded
either as highest- or lowest-weight representations.

\subsection{Generating the $\su(2)_{-1/2}$ spectrum from the vacuum
singular vector}

Another way of generating the spectrum of the model is to use the
constraints induced by the
presence of the non-trivial vacuum singular vector, cf.
\cite{Gab}.\footnote{In \cite{Gab}, this
technique is presented as a simplified implementation of  the Zhu's
algebra. It can be traced back to early works in CFT. For instance,
in \cite{ZZ}, the field associated to the non-trivial vacuum
singular vector is called the {\it model's equation of motion}, and
the way the full spectrum can
be extracted from it is illustrated for the minimal model describing
the Yang-Lee singularity. This approach is also used extensively in the
construction of new $W$ algebras in \cite{Eho}, for example.}
The $\su(2)_{-1/2}$ vacuum representation has a null state at level 4.
It can be written explicitly as\footnote{The singular vectors for the
admissible representations were obtained long ago by
Malikov, Feigin and Fuchs \cite{MFF}, and are referred to as MFF singular
vectors. Their theorem 3.2 \cite{MFF} expresses a singular vector as a
monomial involving fractional powers: for the vacuum representation of the
$k=-1/2$ model, the MFF singular vector reads
$$
    (J^+_{-1})^{7/2} (J^-_0)^{2} (J^+_{-1})^{1/2} |0\R
$$
Using the commutation relations, one can
re-express this result as a sum of terms where each generator has an
integer power. The point here is the following.
The vector (\ref{Nmff}) and the above MFF singular vector both have $L_0$
eigenvalue 4. However, the MFF singular vector has $J_0^3$ eigenvalue 2,
while (\ref{Nmff}) has $J_0^3$ eigenvalue 0. Thus, $J_0^-$ must be
applied twice to get the latter from the former.}
\begin{eqnarray}
   &&\left( J_{-4}^3 - \frac{14}{127} J_{-3}^-J_{-1}^+ +\frac{184}{127}
    J_{-3}^3J_{-1}^3 -
    \frac{62}{127}J_{-3}^+J_{-1}^-+\frac{27}{127}J_{-2}^-J_{-2}^+\right.\nn
   &&\left.-\frac{38}{127}J_{-2}^3J_{-2}^3+
    \frac{100}{127}J_{-2}^-J_{-1}^3J_{1}^++
    \frac{64}{127}J_{-2}^3J_{-1}^-J_{-1}^+-
    \frac{16}{127}J_{-2}^3J_{-1}^3 J_{-1}^3\right. \nn
   &&\left. -\frac{68}{127}J_{-2}^+J_{-1}^-J_{-1}^3
    +\frac{16}{127}J_{-1}^-J_{-1}^-J_{-1}^+J_{-1}^+ -  \frac{16}{127}
    J_{-1}^-J_{-1}^3J_{-1}^3J_{-1}^+
    -\frac{32}{127}J_{-1}^3J_{-1}^3J_{-1}^3J_{-1}^3 \right)\ket{0}
\label{Nmff}
\end{eqnarray}
The zero mode of this null vector provides a constraint that fixes the
allowed representations of the theory. We get the following conditions
\be
\label{casi}
    (3 + 16 C) \left[3 (J_0^3)^2 - C\right]= 0
\ee
where the Casimir, $C$, is given by
\be
    C=\frac{1}{2}( J^-_0J^+_0+J^+_0J^-_0) + J_0^3 J_0^3
\ee
Let us look at these conditions on the  highest-weight state of the
highest-weight representation of spin $j$, denoted
$D_j^+$. We will write the state associated to the field in the
admissible representation of spin $j$ as $|j,m\R_n$, where $m$ and $n$ are
the $J^3_0$ and $L_0$ eigenvalues, respectively:
\begin{eqnarray}
   C|j,m\R_n &=& j(j+1) |j,m\R_n \nn
   J^3_0 |j,m\R_n&=& m |j,m\R_n\nn
   L_0 |j,m\R_n&=&(h_j+n)|j,m\R_n
\end{eqnarray}
The highest-weight state of $D_j^+$ satisfies $J^+_0|j,j\R_0=0$.

{}From the condition (\ref{casi}), we find the following set of
highest-weight representations $D_j^+$:
\begin{eqnarray}
   &&j = 0,\phantom{-}\phantom{/2} \ \ \ \ C=0 \nn
   &&j = 1/2, \phantom{-}\ \ \ \ C=3/4 \nn
   &&j = -1/4, \ \ \ \ C=-3/16 \nn
   &&j = -3/4, \ \ \ \ C=-3/16
\end{eqnarray}
These are in one-to-one correspondence with the admissible
representations for the $\su(2)_{-1/2}$ model.

Also, to each highest-weight representation, there corresponds a lowest-weight
representation, with the same Casimir eigenvalue. The lowest-weight states are
$|j,-j\R_0$, and the corresponding representations are denoted by $D^-_j$.
Note that the states in $D^+_j$ are different from those in $D^-_j$
unless $j$ is integer or half-integer (in which case the corresponding
$su(2)$ representation -- the grade-zero content of
the affine representation -- is finite-dimensional). For
$j=1/2$ (or the trivial $j=0$) the representation is both a
lowest- and a highest-weight representation. The two notions can then
be identified.

As pointed out in \cite{Gab} in a related context, another
infinite-dimensional representation is allowed.
It is neither a highest- nor a
lowest-weight representation, and for that reason it cannot be
assigned a $j$ value. Nevertheless, the representation does have a
well-defined Casimir eigenvalue, $C=-3/16$, and is defined by
\begin{eqnarray}
   J_0^3 |m\R_0 &=& |m+1\R_0\nn
   J_0^+|m\R_0&=&|m+1\R_0\nn
   J_0^- |m\R_0 &=& ( -3/16 - m(m-1) )|m-1\R_0
\end{eqnarray}
It is denoted $E$. In fact, this representation extends to a continuous 
set of representations by setting $m\in\ZZ + t$ with $t\in[0,1[$, excluding
the values of $t$ used for the admissible representations obtained
above (i.e., $t\neq -1/4,-3/4$). These are denoted $E_t$.

In this approach to the determination of the spectrum, it appears
that both highest- and lowest-weight representations are
present. Again, the distinction is meaningful when
$j\not\in \NN/2$, i.e., when $s\neq0$ with $s$ defined in (\ref{rsnot}).
Two such representations always appear pairwise
$(j,-1-j)$, and they have the same conformal dimension.
Moreover, the associated character function $\chi^+_j$, viewed as a
highest-weight representation (hence the superscript $+$), is related to the
conjugate character function $\chi^-_{-1-j} $, viewed as a
lowest-weight representations \cite{Ramgoolam, Imbimbo}:
\be
\label{conj}
    \chi^+_j(\tau,z)= -\chi^-_{-1-j}(\tau,-z)\ \ \ \ \ \ \ (s\neq0)
\ee

For later use, we mention that in order
to describe the primary fields associated to representations which are
infinite-dimensional at grade zero, it is convenient to use a collective
description of the whole multiplet in terms of a generating function
\cite{ZF}. The expansion in powers of the dummy variable $x$ captures the
$m$ values of the multiplet. For instance, to a highest-weight
representation with $j\not\in\NN/2$, there corresponds the field
\be
     \phi_j^+(w,x)= \sum_{m\in(j+\ZZ_\leq)}  \phi^{(m)}_j(w) \,x^{j+m}
\label{primp}
\ee
while the field associated to a lowest-weight representation reads
\be
     \phi_j^-(w,x)= \sum_{m\in(-j+\ZZ_\geq)}  \phi^{(m)}_j(w) \,x^{j+m}
\label{primm}
\ee

Having this set of highest- and lowest-weight representations,
the question arises: which representations are
actually present in the Kac-Wakimoto diagonal modular invariant?
When revisiting the
diagonal invariant (\ref{Ze}), it is not difficult to realize that it
could just as well be written either as
\be
   Z=|\chi_{0}(\tau,z)|^2+|\chi_{1/2}(\tau,z)|^2+|\chi_{-1/4}^+(\tau,z)|^2+
    |\chi_{-1/4}^-(\tau,-z)|^2
\ee
or
\be
   Z=|\chi_{0}(\tau,z)|^2 +|\chi_{1/2}(\tau,z)|^2
    +|\chi_{-3/4}^+(\tau,z)|^2+|\chi_{-3/4}^-(\tau,-z)|^2
\ee
This would seem to indicate a contradiction:
either there are more than one theory for which the operator product algebra
closes, or we do not have a correct interpretation
of the partition function. It turns out that the latter is the
correct answer, and that we need to worry about convergence regions
when talking about characters. To elucidate this, we must consider products
of operators to determine which set of operators closes under fusion.
This will not be tackled fully until Section 6. The next subsection
is devoted to an examination of the standard approaches
to computing fusion rules in the $\su(2)_{-1/2}$ WZW model.

\subsection{Fusion rules}

In the context of non-unitary WZW models, there  are contradicting
results concerning fusion rules. The method of BRST cohomology
\cite{BF} and the vertex-operator methods \cite{DLM} yield
identical results. On the
other hand, fusion rules can also be computed by
enforcing the decoupling of singular vectors \cite{AY}. The two
sets of results are rather different. The latter fusion rules
were recovered in \cite{FM} using cohomology theory of
infinite-dimensional Lie algebras,
and in \cite{PRY} using a Coulomb gas method based on the Wakimoto free-field
realization to compute correlators. Notice also that the Verlinde formula
incorporates the results of \cite{BF}, but in those cases where the
latter are ill-defined, it yields negative signs \cite{MWa}.

As a preparation for our discussions in subsequent sections,
we here review both sets of results for $k=-1/2$. In
both cases, we omit discussing the trivial fusions with the identity
field. First, the fusion rules
obtained by decoupling the singular vectors \cite{AY} are
\begin{eqnarray}
\label{ayfu}
    D^+_{-3/4}\times D^-_{-3/4}&=&{\phantom{-}}D_0\nn
    D^+_{-1/4} \times D^-_{-1/4} &=&{\phantom{-}}D_0\nn
    D^\pm_{-3/4} \times D^\mp_{-1/4}&=&{\phantom{-}}D_{1/2}\nn
    D^\pm_{-3/4}\times D_{1/2\phantom{-}}&=&{\phantom{-}}D^\pm_{-1/4}\nn
    D^\pm_{-1/4}\times D_{1/2\phantom{-}}&=&{\phantom{-}}D^\pm_{-3/4}\nn
    D_{1/2\phantom{-}}\times D_{1/2\phantom{-}}&=&{\phantom{-}}D_{0}
\end{eqnarray}
One observes that all fusions combine a highest- with a
lowest-weight representation: $D^+\times D^-$ or $D^-\times D^+$.
Recall that $D_0$ and $D_{1/2}$ are both highest- and lowest-weight
representations.

Let us explain the reason for which we insist on interpreting these fusion
rules in terms of fields with specified representations.
The derivation presented in \cite{AY} considers the fusion of three
fields $\phi_j(z,x)$. Two of the associated representations
(highest, lowest or even continuous) are initially not specified,
while one field carries a highest-weight representation
(or equivalently a lowest-weight representation).
Assuming it is the field $\phi_{j_1}(z_1,x_1)$, one then considers
the decoupling of the singular vector of the highest-weight representation
$j_1$ from the three-point function
\be
\L j_3|\phi_{j_2}(z_2,x_2)|j_1\R
\ee
Applying the null vector to this correlator then provides a set of
products depending on
$j_1,j_2,j_3$ which must be set to zero in order to have decoupling.
This leads to a set of conditions on the three spins, thus
providing the fusion rules. At first sight, it then seems that
only the representation of the field $\phi_{j_1}(z_1,x_1)$ has been
specified. However that is not true. From projective and global
$SU(2)$ invariance, one obtains the well-known form
of the generating-function three-point correlator:
\ben
   \L\phi_{j_3}(z_3,x_3)\phi_{j_2}(z_2,x_2)\phi_{j_1}(z_1,x_1)\R
    \propto \frac{(x_2-x_1)^{j_2+j_1-j_3}(x_3-x_2)^{j_3+j_2-j_1}
     (x_3-x_1)^{j_3+j_1-j_2}}{(z_2-z_1)^{h_2+h_1-h_3}(z_3-z_2)^{h_3+h_2-h_1}
     (z_3-z_1)^{h_3+h_1-h_2}}
\een
One observes that for the triplets of spins appearing in the
fusion rules (\ref{ayfu}), at most one of the combinations
$j_2+j_1-j_3$, $j_3+j_2-j_1$ and $j_3+j_1-j_2$ is not a non-negative integer.
This means that at most one of the monomials ($(x_2-x_1)^{j_2+j_1-j_3}$, say)
requires an infinite expansion. As a result, one of the involved
fields will correspond to an infinite-dimensional
highest-weight representation, while the other will correspond
to an infinite-dimensional lowest-weight representation.
This is reflected in the fusion rules. At most two
of the representations are infinite-dimensional, and in the cases
where they appear on each side of the fusion identity,
they are both highest or lowest weight.
That is due to the fact that three-point functions correspond
to couplings of three representations to the singlet.
Thus, extracting fusion rules from three-point functions requires
considering the conjugate representation to one of the fields, interchanging
a highest- with a lowest-weight representations.


In a rational CFT, the fusion algebra of the finitely many primary fields
must close.  However, the decoupling method does not predict the outcome
of the fusion $D^+_{-1/4} \times D^+_{-1/4}$, for example.
One may also worry about the associated modular invariant.
In that vein, it is noted that the fusion rules (\ref{ayfu})
seem to superficially contradict the various forms of the modular
invariant written before, none of which, at first sight,
contains all the fields (associated to $D_0$, $D_{1/2}$,
$D^\pm_{-1/4}$ and $D^\pm_{-3/4}$) at the same time.

Note finally that the fusions (\ref{ayfu}) are invariant under the action
of the $\su(2)$ outer automorphism $a$ which acts on the spin labels as
$a: j\rightarrow a(j)=k/2-j$. The
invariance property of the fusion rules is
\be
\label{auto}
D_{a(j_1)}\times D_{a'(j_2)}= \sum_{j_3} D_{aa'(j_3)}
\ee
with the $+/-$ specification omitted.
A more refined version of this symmetry relation will be considered later.

Another set of fusion rules are computed from a direct
application of the Verlinde formula. In the case of the
diagonal modular invariant containing all highest-weight representations,
the rules are (cf. \cite{MWa}):
\begin{eqnarray}
\label{verli}
D^+_{-3/4}\times D^+_{-3/4}&=&-D_{1/2} \nn
D^+_{-1/4} \times D^+_{-1/4}&=&-D_{1/2} \nn
D^+_{-3/4} \times D^+_{-1/4}&=&-D_{0\phantom{/2}} \nn
D^+_{-3/4}\times D_{1/2\phantom{-}}&=&{\phantom{-}}D^+_{-1/4}\nn
D^+_{-1/4}\times D_{1/2\phantom{-}}&=&{\phantom{-}}D^+_{-3/4}\nn
D_{1/2\phantom{-}}\times D_{1/2\phantom{-}}&=&{\phantom{-}}D_{0}
\end{eqnarray}
(here the representations $D_0$ and $D_{1/2}$ are viewed as
highest-weight representations). The Bernard-Felder rules \cite{BF}
correspond to considering only the last three fusions.

The two sets of fusion rules are obviously different, though they do
have common features. In particular, all fusions with the identity
(not written) as well as the last three fusions in each set,
are identical (when restricting to highest-weight representations).
However, the first three are rather different. These
are precisely the ones that involve a negative sign in the Verlinde
case. An immediate consequence is that  the Verlinde fusions are not
invariant under (\ref{auto}) (with all highest weights).
However, the relation (\ref{conj}) suggests a simple way to re-conciliate
these different results \cite{Ramgoolam, Imbimbo},
and to interpret the negative signs, e.g.,
\be
D^+_{-3/4}\times D^+_{-3/4}=-D_{1/2}
\qquad \rightarrow \qquad D^+_{-3/4}\times(-D^-_{-1/4})=-D_{1/2}
\ee
Although appealing at first sight, we stress that this
prescription cannot make the
two sets of fusion rules identical in the general case (that is, when
$u\geq 3$, where $u$ is the denominator of
$k$) since the decoupling method yields more terms than predicted by
the Verlinde formula. Roughly, the fusion rules in both methods split
into separated fusions in the integral
and fractional sector, that is, in terms of the $r$ and $s$ labels
(\ref{rsnot}). The decoupling fusions are isomorphic to those of
$\su(2)_{u(k+2)-2}\otimes {\widehat {osp}}(1,2)_{u-1}$, while the
fusions obtained by the Verlinde formula
are of the type $\su(2)_{u(k+2)-2}\otimes \widehat{u}(1)_{u-1}$
(see, e.g., \cite{MRW}).

The apparent contradiction between the two sets of fusion rules
is a further signal that the CFT interpretation of the WZW model with
only four primary fields may be wrong.
In order to address in a concrete way the issues raised here, we first
describe the connection between the $\su(2)_{-1/2}$ WZW model and the
$\bg$ system.
That will provide us with a free-field representation of the WZW model.
It is established in the next section, and confirmed in Section
5 at the level of correlation functions. Thus armed,
we revisit in Section 6
the various puzzling issues raised in this section.

\section{The $\su(2)_{-1/2}$ model vs the $\bg $ system}

\subsection{The $\bg$ representation of the current algebra}

As pointed out in \cite{LudwigNPB519}, the $\su(2)$ currents live in
the universal covering of the $\bg$ algebra. More precisely, each current
can be represented as a
bilinear in these ghosts as
\begin{eqnarray} \label{curre}
   J^+ & = & \phantom{-}\frac{1}{2} :\beta^2: \nn
   J^- &  = &-\frac{1}{2}:\gamma^2 : \nn
   J^3 &  = & -\frac{1}{2} :\gamma \beta :
\end{eqnarray}
The OPEs read
\begin{eqnarray}
   J^+(z)J^-(w)& \sim & {-1/2\over (z-w)^2}+{2J^3(w)\over (z-w)}\nn
   J^3(z)J^\pm(w)& \sim & {\pm J^\pm(w)\over (z-w)}\nn
   J^3(z)J^3(w)& \sim & {-1/4\over (z-w)^2}
\end{eqnarray}

Having found the representation of the currents, the  next step is to
understand the structure of the WZW primary fields in terms of the
$\beta\gamma$ system. It is rather easy to verify that the $(\beta,\gamma)$
pair forms a spin-$1/2$ multiplet:
\bea
   &&J^3(z)\beta(w) \ \sim \  \phantom{-}\frac{1}{2}{\beta(w)\over
    (z-w)},\qquad J^+(z)\beta(w)\ \sim \ 0\nn
   &&J^3(z)\gamma(w )\ \sim \ -\frac{1}{2}{\gamma(w)\over (z-w)},\qquad
    J^-(z)\beta(w) \ \sim \ 0\nn
   &&J^-(z)\beta(w) \ \sim\
    {\gamma(w)\over (z-w)}, \qquad  J^+(z)\gamma(w)\ \sim\
    {\beta(w)\over (z-w)}
\eea
This means that $\beta$ and $\gamma$ correspond to the states
$|1/2,1/2\R_0$ and $|1/2,-1/2\R_0$, respectively.

The crux of the matter is to describe the operators with weight
$h=-{1\over 8}$. They cannot  be expressed directly in terms of
the
$\bg$ fields. The natural guess is to identify them with the
$\ZZ_2$ twists in the $\bg$ system. We then need to understand how the
twist fields are organized to form irreducible representations.
In particular, we must be able to describe the infinite number of
states at grade zero in the $j=-1/4$ and $j=-3/4$ admissible
representations. To proceed further along these lines, we turn to the
$\eta\xi\phi$ free-field representation.

\subsection{Twist fields and representations of  $\su(2)_{-1/2}$}

In terms of the $\eta\xi\phi$ fields,
the $\su(2)_{-1/2}$ currents are represented by
\begin{eqnarray}
   J^{+}&=&{1\over 2}e^{-2i\phi}:\partial\eta\eta:\nn
    J^{-}&=&{1\over 2}e^{2i\phi}:\partial^{2}\xi\partial\xi:\nn
    J^{3}&=&{1\over 2}i\partial\phi
\label{Jferm}
\end{eqnarray}
We now study their action on the twist fields, noting that
\be
    J^{3}_0\tau_{\lambda}= -{(2\lambda-1)\over 4}\tau_{\lambda}
\ee
Let us first identify those twist fields that correspond to highest-
and lowest-weight states (characterized by the vanishing of the
single pole in the OPE with $J^\pm$, respectively).
There are two highest weights (hw) and two lowest weights (lw):
\begin{eqnarray}
    J^-_0\tau_{0}=0,\ ~~J^{3}_0\tau_{0\phantom{-}}=\phantom{-}{1\over
     4}\tau_{0\phantom{-}}\quad
     &\Rightarrow&\quad {\rm
     lw:}\; j=-{1\over 4},\; m=\phantom{-}{1\over 4}\nn
     J^-_0\tau_{-1}=0,\ ~~J^{3}_0\tau_{-1}=\phantom{-}{3\over 4}\tau_{-1}\quad
     &\Rightarrow&\quad {\rm
     lw:}\; j=-{3\over 4},\; m=\phantom{-}{3\over 4}\nn
     J^+_0\tau_{1}=0,\ ~~J^{3}_0\tau_{1\phantom{-}}=-{1\over
     4}\tau_{1\phantom{-}}\quad
     &\Rightarrow&\quad {\rm
     hw:}\; j=-{1\over 4},\; m=-{1\over 4}\nn
    J^+_0\tau_{2}=0,\ ~~J^{3}_0\tau_{2\phantom{-}}=-{3\over
    4}\tau_{2\phantom{-}}\quad
    &\Rightarrow&\quad {\rm
    hw:}\; j=-{3\over 4},\; m=-{3\over 4}
\end{eqnarray}
More generally, the twist fields with $\la\in\ZZ$ can be organized to form
the admissible representations:
\begin{eqnarray}
       \tau_{-2n},~~n\in \ZZ_\geq:\ \ ~~D_{-1/4}^{-}\nonumber\\
       \tau_{-2n-1},~~n\in \ZZ_\geq:\ \ ~~D_{-3/4}^{-}\nonumber\\
       \tau_{2n+1},~~n\in \ZZ_\geq:\ \ ~~D_{-1/4}^{+}\nonumber\\
       \tau_{2n+2},~~n\in \ZZ_\geq:\ \ ~~D_{-3/4}^{+}
\end{eqnarray}
These twist fields are all expressed in terms of the free boson and
   differential monomials in $\eta$ or $\xi$. Moreover,
the above result indicates that
generic twist fields are in representations which are neither highest-
nor lowest-weight representations!
Indeed, for $\la\not\in\ZZ$, the twist fields
$\tau_{\lambda+2n}$ ($n\in \ZZ$) form infinite-dimensional
representations which are neither highest- nor lowest-weight representations.

We consider the $\bg$ free-field representation of the $\su(2)_{-1/2}$
model further by examining four-point functions in the next section.

\section{The KZ equation and correlators}

In this section we want to construct the four-point functions for the
$\su(2)_{-1/2}$ model using the KZ equation. Let us
introduce the differential operator realization of the $su(2)$ generators
\be
     J^+_0 = -x^2\partial_x + 2 j x, \ \ \ \
     J^-_0=\partial_x, \ \ \ \ J^3_0 = x\partial_x - j
\label{diffo}
\ee
The Casimir operator is given by
\be
     C = \eta_{ab} J^a_0 J^b_0 = j (j+1)
\ee
with $\eta_{+-}=\eta_{-+}= \eta_{33}/2=1/2$.
Note that the WZW primary fields $\phi_j(w,x)$ satisfy
\be
     J^a(z)\phi_j(w,x)\ \sim \ \frac{1}{z-w} J^a_0\phi_j(w,x)
\ee

We recall that the KZ equation captures the
constraint that follows from the insertion of the null vector
\be
   \left(L_{-1}-\frac{1}{k+2}\eta_{ab} J_{-1}^a J_0^b\right)|\phi\R
\ee
into a four-point function, for example. Using the realization (\ref{diffo}),
the four-point KZ equation reads
\be
    \left[ (k+2)\partial_{z_i} + \sum_{j\neq i}\frac{\eta_{ab} (J_0^a)_i 
 \otimes (J_0^b)_j}{z_i-z_j}\right]
    \L\phi_{j_4}(z_4,x_4)\phi_{j_3}(z_3,x_3)\phi_{j_2}(z_2,x_2)
    \phi_{j_1}(z_1,x_1)\R = 0
\ee
Here $(J_0^a)_j$ means that $J^a_0$ acts on the field labeled by
$j$ and positioned at $z_j$. The generic solution to these equations
was computed in \cite{Nichols}, and contains logarithms.

{}From the similarity with the $c=2$ CFT, for example, it is natural to
expect that a solution to the KZ equation may involve an infinite
series in $x_i$, and not just be polynomial.
This is indeed what happens. Introduce the differential operators
\begin{eqnarray}
    P&=&-x^2(1-x)\partial^2_x+\left[-(j_1+j_2+j_3-j_4+1)x^2+2j_1x+2j_2x(1-x)
     \right]\partial_x \nn
    &&+2j_2(j_1+j_2+j_3-j_4)x-2j_1j_2
\end{eqnarray}
and
\bea
    Q&=&-(1-x)^2x\partial^2_x+
     \left[(j_1+j_2+j_3-j_4+1)(1-x)^2+2j_3(1-x)\right.\nn
    &&\left.+2j_2x(1-x)\right]\partial_x
     +2j_2(j_1+j_2+j_3-j_4)(1-x)-2j_2j_3
\eea
The anharmonic ratios, $z$ and $x$, are defined by
\ben
     z=\frac{(z_2-z_1)(z_4-z_3)}{(z_3-z_1)(z_4-z_2)},\ \ \ \
     x=\frac{(x_2-x_1)(x_4-x_3)}{(x_3-x_1)(x_4-x_2)}
\label{anhx}
\een
The KZ equation may then be written
\be
    (k+2)\partial_z F_{j_1,j_2,j_3,j_4}(z,x) = \left[
    \frac{P}{z}+\frac{Q}{z-1}\right]F_{j_1,j_2,j_3,j_4}(z,x)
\ee
where the function $F_{j_1,j_2,j_3,j_4}$ is defined by
\be
    F(z,x)_{j_1,j_2,j_3,j_4}=\lim_{z_4,x_4\rightarrow \infty}
    z_4^{2h_{j_4}}x_4^{-2j_4}\L\phi_{j_4}(z_4,x_4)\phi_{j_3}(1,1)
    \phi_{j_2}(z,x)\phi_{j_1}(0,0)\R
\label{defF}
\ee

Some solutions to the KZ equation are related to each other under
the action of the outer automorphism $a$. Indeed, under the map
$j_i\rightarrow \frac{k}{2}-j_i$,
$F_{j_1,j_2,j_3,j_4}$ may be shown to obey the same KZ equation as
\be
    (x-z)^{k-j_1-j_2-j_3-j_4} z^\alpha (1-z)^\beta
    F_{\frac{k}{2}-j_3, \frac{k}{2}-j_4, \frac{k}{2}-j_2, \frac{k}{2}-j_1}
\label{Fji}
\ee
$\alpha$ and $\beta$ are determined through
the commutation of (\ref{Fji}) with the differential operators $P$ and $Q$.
The correlators of interest here are the ones for $k=-1/2$ involving fields
with dimension $h=-1/8$. Let us focus on the case $j_1=j_2=j_3=j_4=-1/4$.
Relating the solution \cite{Nichols}
\be
    F(z,x)_{0,0,0,0} = A'\left[ \frac{2\log z}{3} +\log x\right] +
     B' \left[ \frac{2\log (1-z)}{3} +\log(1-x) \right] + C'
\ee
(where $A',B'$ and $C'$ are constants) to $F(z,x)_{-1/4,-1/4,-1/4,-1/4}$,
we find $\al=\beta=1/4$, and hence
\bea
    &&F(z,x)_{-1/4,-1/4,-1/4,-1/4}
     \ =\ \frac{[z(1-z)]^{\frac{1}{4}}}{\sqrt{(x-z)}}\nn
    &&\times\left\{ A' \left[   \frac{2\log z}{3} +\log x\right]
     +B'\left[ \frac{2\log (1-z)}{3} +\log(1-x)\right] + C' \right\}
\label{horrid}
\label{F4}
\eea

Other solutions may be obtained using a symmetry under $j\rightarrow-1-j$.
The symmetry is expressed as
\be
   \frac{\partial^{1+2j_2}}{\partial x^{1+2j_2}} F_{j_1, j_2, j_3,
    j_4}(z,x) = F_{j_1, -1-j_2, j_3, j_4}(z,x)
\ee
As discussed in \cite{Nichols}, this equation only makes sense
when $1+2j \in \ZZ_\geq$. However, if we
consider the change of more than
one field, we can formally combine fractional derivatives to obtain
a well-defined operation. This is useful here since the spins are
$j=-1/4$, so that $1+2 j = 1/2$. Thus, the application of this operation
on two spins relates, by a simple derivative, the solution (\ref{F4}) to
one for which two of the spins are now $j=-3/4$.

\subsection{Generating function for twist correlators}

We will postpone the discussion of logarithms to our subsequent paper
\cite{LMRS},
and here concentrate on the simplest solution deduced from (\ref{horrid}):
\ben
     z^{1/4}(1-z)^{1/4}(x_3-x_1)^{-1/2}(x_4-x_2)^{-1/2}
      \left(\frac{(x_1-x_2)(x_3-x_4)}{(x_3-x_1)(x_4-x_2)}-z\right)^{-1/2}
\label{F4a}
\een
All the prefactors left over in the study of the KZ equation have been
re-installed. Our goal is to prove that this expression is a
generating function for the four-point functions of twist correlators
determined in Section 2, hence providing further support to our
interpretation of the $\beta\gamma$ system as an $\su(2)_{-1/2}$ theory.

In order to be able to expand (\ref{F4a}) we choose to consider the
region defined by
\ben
     z<x,\ \ \ \ x_1<x_2<x_3<x_4
\label{order}
\een
This is also the ``natural region'' to consider from the point
of view of using loop projective invariance to fix three of the $x_i$'s
to the standard values, cf. (\ref{defF}).
We may now expand the $x$-dependent part of (\ref{F4a}):
\bea
     &&(x_3-x_1)^{-1/2}(x_4-x_2)^{-1/2}
      \left(\frac{(x_2-x_1)(x_4-x_3)}{(x_3-x_1)(x_4-x_2)}-z\right)^{-1/2}\nn
     &=&(x_2-x_1)^{-1/2}(x_4-x_3)^{-1/2}\sum_{n\geq0}
      \bin{-1/2}{n}
      \left(-z\frac{(x_3-x_1)(x_4-x_2)}{(x_2-x_1)(x_4-x_3)}\right)^n\nn
     &=&(x_2x_4)^{-1/2}\sum_{n,m_1,m_2,m_3,m_4\geq0}(-1)^{n+m_1+m_2+m_3+m_4}
      \bin{-1/2}{n}z^n\nn
     &\times&\bin{-1/2-n}{m_1}\bin{-1/2-n}{m_2}\bin{n}{m_3}\bin{n}{m_4}
       x_1^{N_1}x_2^{N_2}x_3^{N_3}x_4^{N_4}
\label{F4b}
\eea
where we have introduced
\bea
     N_1&=&m_1+m_3\nn
     N_2&=&-m_1+m_4-n\nn
     N_3&=&m_2-m_3+n\nn
     N_4&=&-m_2-m_4
\label{N}
\eea
A first and almost trivial observation is that
\ben
     N_1+N_2+N_3+N_4=0
\label{NNNN0}
\een
which simply corresponds to the conservation of momentum (and follows from
global $SU(2)$ invariance).
The coefficient to $x_1^{N_1}x_2^{N_2-1/2}x_3^{N_3}x_4^{N_4-1/2}$ may now
be evaluated, and using (\ref{N}) we find
\bea
   X(N_1,N_2,N_3,N_4;z)
    &=&\sum_{n\geq0}(-1)^{N_2+N_3+n}\bin{-1/2}{n}z^n\sum_{m\geq0}
     \bin{-1/2-n}{m}\nn
    &\times&\bin{-1/2-n}{N_1+N_3-n-m}\bin{n}{N_1-m}\bin{n}{N_2+n+m}
\label{binomials}
\eea
One has to be careful when analyzing this double summation.
The summation ranges of $n$ and $m$ must be cut, and one is left with
a sum of several double summations. In each of them, one may
express the summation over $m$ as a terminating ${}_4F_3$
hypergeometric function with argument 1.
Unfortunately, the hypergeometric functions are not balanced,
while most known results pertain to such functions.
In particular, the 6-$j$ symbols are associated to balanced
hypergeometric functions.

Nevertheless, according to the twist-field approach,
we should expect to be able
to express (\ref{binomials}) in terms of an ordinary hypergeometric
function ${}_2F_1(a,b;c;z)$ with $a,b,c$ depending on $N_1,N_2,N_3$.
Indeed, we find
\bea
    X(N_1,N_2,N_3,N_4;z)
    &=&(-1)^{N_1+N_3}\bin{-1/2}{N_1}
     \bin{-1/2+N_3+N_4}{N_3}\bin{N_1}{N_1+N_2}\nn
    &\times&z^{N_1+N_2}(1-z)^{-(N_2+N_3)}{}_2F_1(-2N_3,1+2N_1;
     1+2|N_1+N_2|;z)\nn
    &+&(-1)^{N_1+N_3}\bin{-1/2}{N_3}
     \bin{-1/2+N_1+N_2}{N_1}\bin{N_3}{N_3+N_4}\nn
    &\times&z^{N_3+N_4}(1-z)^{-(N_1+N_4)}{}_2F_1(-2N_1,1+2N_3;
     1+2|N_3+N_4|;z)\nn
    &-&\delta_{N_1+N_2,0}(-1)^{N_1+N_3}
     \bin{-1/2}{N_1}\bin{-1/2}{N_3}\nn
    &\times&(1-z)^{N_1-N_3}{}_2F_1(-2N_3,1+2N_1;1;z)
\label{Xnnn}
\eea
$N_4$ has been included to emphasize the symmetry
\begin{equation}
    X(N_1,N_2,N_3,N_4;z)=X(N_3,N_4,N_1,N_2;z)
\end{equation}
Note that
\be
   (1-z)^{N_1-N_3}{}_2F_1(-2N_3,1+2N_1;1;z)
    =(1-z)^{-N_1+N_3}{}_2F_1(-2N_1,1+2N_3;1;z)
\ee
showing the symmetry of the final term (\ref{Xnnn}).
The absolute values ensure that the
hypergeometric functions are well-defined, while the binomials
split the result according to $N_1+N_2\geq0$ or $N_1+N_2\leq0$.
The overlap $N_1+N_2=0$ is taken care of by the final subtraction.

We have verified (\ref{Xnnn}) in many explicit examples,
and been able to prove it analytically in several cases.
One of those cases is the
particularly interesting situation when $N_1=-N_2=N_3=-N_4$:
\ben
     X(N,-N,N,-N;z)=\bin{-1/2}{N}\bin{-1/2}{N}{}_2F_1(2N+1,-2N;1;z)
\label{NNN}
\een
Let us indicate how one may prove this result. We see that
(\ref{binomials}) reduces to
\bea
   X(N,-N,N,-N;z)&=&\left(\sum_{n=0}^N\ \sum_{m=N-n}^N+
    \sum_{n=N+1}^{2N}\ \sum_{m=0}^{2N-n}\right)(-1)^nz^n\nn
   &\times&\bin{-1/2}{n}\bin{-1/2-n}{m}\bin{-1/2-n}{2N-n-m}
     \bin{n}{N-m}\bin{n}{N-m}\nn
   &=&\sum_{n=0}^N(-1)^nz^n\bin{-1/2}{n}\bin{-1/2-n}{N-n}\bin{-1/2-n}{N}\nn
   &\times&{}_4F_3\left[\begin{array}{lll}N+1/2, \ -N, \ -n, \ -n\\
    N-n+1, \ -N-n+1/2, \ 1 \end{array};\ 1\right]\nn
   &+&\sum_{n=N+1}^{2N}(-1)^nz^n\bin{-1/2}{n}\bin{-1/2-n}{2N-n}
     \bin{n}{N}\bin{n}{N}\nn
   &\times&{}_4F_3\left[\begin{array}{lll}n+1/2, \ -N, \ -N, \ -(2N-n)\\
    -2N+1/2, \ n-N+1, \ n-N+1 \end{array};\ 1\right]
\label{sumX}
\eea
Now we use Eq. (2.4.2.3) of \cite{Slater}:
\bea
   &&{}_4F_3\left[\begin{array}{lll}f,\ 1+f-h,\ h-a,\ d\\
    h,\ 1+f+a-h,\ g \end{array};\ 1\right]\nn
   &=&\frac{\Gamma(g)\Gamma(g-f-d)}{
     \Gamma(g-f)\Gamma(g-d)}
    \ {}_5F_4\left[\begin{array}{lll}a,\ d,\ 1+f-g,\ \frac{1}{2}f,\
    \frac{1}{2}(1+f)\\
     h,\ 1+f+a-h,\ \frac{1}{2}(1+f+d-g),\ \frac{1}{2}(2+f+d-g) \end{array}
    ;\ 1\right]
\label{Slater}
\eea
which is valid when $f$ or $d$ is a negative integer.
Applying (\ref{Slater}) to the two hypergeometric functions of our
interest (\ref{sumX}), the resulting ${}_5F_4$ hypergeometric
functions reduce to Saalschutzian ${}_3F_2$ hypergeometric functions.
They may be summed using Saalschutz's theorem, and (\ref{NNN}) follows
straightforwardly.

Other interesting situations appear when one of the $N_i$ vanishes.
It is straightforward to verify analytically that (\ref{binomials})
then sums to
\bea
     &&X(0,N_2,N_3,N_4;z)
       =(-1)^{N_3}\bin{-1/2}{-N_2}\bin{-1/2+N_2}{-N_4}z^{-N_2}
      (1-z)^{-N_4}\nn
     &&X(N_1,0,N_3,N_4;z)
        =(-1)^{N_4}\bin{-1/2}{N_1}\bin{-1/2-N_1}{N_3}z^{N_1}
      (1-z)^{N_3}\nn
     &&X(N_1,N_2,0,N_4;z)=(-1)^{N_1}\bin{-1/2}{-N_4}\bin{-1/2+N_4}{-N_2}
      z^{-N_4}(1-z)^{-N_2}\nn
     &&X(N_1,N_2,N_3,0;z)=(-1)^{N_2}\bin{-1/2}{N_3}\bin{-1/2-N_3}{N_1}
      z^{N_3}(1-z)^{N_1}
\label{N0}
\eea
with the arguments still subject to (\ref{NNNN0}).

In conclusion, we have found that the general four-point chiral
block is given by
\begin{eqnarray}
    &&\langle\phi_{-1/4}(z_4,x_4)\phi_{-1/4}(z_3,x_3)\phi_{-1/4}(z_2,x_2)
     \phi_{-1/4}(z_1,x_1)\rangle\nn
    &=&(z_3-z_1)^{1/4}(z_4-z_2)^{1/4}z^{1/4}(1-z)^{1/4}\nn
    &\times&\sum_{n_1,n_2,n_3,n_4\geq0}
     x_1^{n_1}x_2^{-1/2-n_2}x_3^{n_3}x_4^{-1/2-n_4}
     (-1)^{n_1+n_3}\delta_{n_1-n_2+n_3-n_4,0}\nn
    &\times&\left\{\bin{-1/2}{n_1}
     \bin{-1/2+n_3-n_4}{n_3}\bin{n_1}{n_1-n_2}\right.\nn
    &\times&z^{n_1-n_2}(1-z)^{n_2-n_3}{}_2F_1\left[-2n_3,1+2n_1;
     1+2|n_1-n_2|;z\right]\nn
    &+&\bin{-1/2}{n_3}
     \bin{-1/2+n_1-n_2}{n_1}\bin{n_3}{n_3-n_4}\nn
    &\times&z^{n_3-n_4}(1-z)^{n_4-n_1}{}_2F_1(-2n_1,1+2n_3;
     1+2|n_3-n_4|;z)\nn
    &-&\left.\delta_{n_1,n_2}\bin{-1/2}{n_1}
     \bin{-1/2}{n_3}(1-z)^{n_1-n_3}{}_2F_1(-2n_3,1+2n_1;1;z)\right\}
     \label{big}
\end{eqnarray}
A hitherto implicit $z$-dependent prefactor has been included,
and the notation has been changed according to $N_1,N_2,N_3,N_4
\rightarrow n_1,-n_2,n_3,-n_4$.

We can now compare this result with the correlators for twist fields
in the $\beta\gamma$ system, recalling the expansions
(\ref{primp}) and (\ref{primm}).
Now, suppose we insert at $z_{1}$ and $z_{3}$ lowest-weight representations
$D_{-1/4}^{-}$, and at $z_{2}$ and $z_{4}$ highest-weight representations
$D_{-1/4}^{+}$. The expansion of the correlator should then read
\be
       x_{2}^{-1/4}x_{4}^{-1/4}\sum_{n_{i}\geq 0}
       x_{1}^{n_{1}} x_{2}^{-n_{2}} x_{3}^{n_{3}} x_{4}^{-n_{4}}
       \L\tau_{1+2n_{4}}\tau_{-2n_{3}}\tau_{1+2n_{2}}\tau_{-2n_1}\R
\ee
A comparison is now straightforward. For $n_{1}-n_{2}\geq0$, the
matching of the first term in (\ref{big}) with the corresponding
twist correlator (\ref{twistcorr}) is obvious. For $n_{1}-n_{2}<0$, we recall that
our simplified prescription for the twist correlators involved
extracting the (possible) pole from the hypergeometric function
by a formal manipulation. This may be achieved using
\bea
    {}_2F_1(a,b;-N;z)&\propto& z^{N+1}{}_2F_1(a+N+1,b+N+1;N+2;z)\nn
    &=&z^{N+1}(1-z)^{-a-b-N}{}_2F_1(-a+1,-b+1;N+2;z)
\eea
and a match with the second term in (\ref{big}) is obtained.

We observe that the choice of order used here (\ref{order})
dictates the representations carried by the primary fields.
That merely reflects
that the generating-function correlator expands on different pairings
of representations depending on the order of its arguments.
An interpretation of the expansion of the primary fields themselves
was addressed in \cite{PRY} in the context of a Wakimoto free-field
realization. There it was argued that the expansions depend
on the numbers of screening operators associated to the various
intertwining operators. It was referred to as adjustable
monodromy, since the primary fields would appear with different
monodromy properties depending on the context. Our results here on the
generating functions are in accordance with \cite{PRY,FGP}.
The first systematic work on correlators in fractional-level
WZW models appeared in \cite{FGPP}.

\section{Fusion rules, characters and the modular invariant}

In this section, we probe the darker corners of our models in order to
address the puzzles raised in the
previous sections.  Our key tool is the free-field representation.
We revisit the
fusion rules, the operator content, and the modular invariant.

\subsection{Twist fusion rules from the free-field representation}

We first return to the $\beta\gamma$ system and use its  $\eta\xi \phi$
representation  to reconsider the
fusion rules, in particular those involving the twist fields.
At first, let us look at the simple fusions involving one twist
field and one ghost field. For instance, we have
\be
    \beta(z)\tau_\la(w)\sim {\tau_{\la-1}(w)\over (z-w)^{1/2}}
\ee
Simple special cases are
\be
    \beta(z)\tau_0(w)\sim {\tau_{-1}(w)\over (z-w)^{1/2}},\
    \qquad \beta(z)\tau_2(w)\sim{\tau_{1}(w)\over (z-w)^{1/2}}
\ee
They are of the types
$D^-_{-1/4}\times D_{1/2}=D^-_{-3/4}$ and $D^+_{-3/4}\times
D_{1/2}=D^+_{-1/4}$, respectively. This result is confirmed by
computations with generic descendants.
Other examples of this type lead to the rules
\begin{eqnarray}
\label{ayfuu}
    D^\pm_{-3/4}\times D_{1/2\phantom{-}}&=&{\phantom{-}}D^\pm_{-1/4}\\
    D^\pm_{-1/4}\times D_{1/2\phantom{-}}&=&{\phantom{-}}D^\pm_{-3/4}
\end{eqnarray}
The $D_{1/2}$ representation intertwines the two infinite-dimensional
representations with $j=-1/4$ and $j=-3/4$. The remaining fusion involving the
$j=1/2$ representation, namely $D_{1/2}\times
D_{1/2}=D_{0}$, is easily verified. These fusions rules agree with
the last three in (\ref{ayfu}) and (\ref{verli}).

Let us now consider fusions involving two twist fields. Take for
instance fusions of the type
$ D^+_{-1/4}\times D_{-1/4}^- $. The simplest case is
\be
    \tau_1(z)\tau_0(w)= e^{i\phi(z)/2}e^{-i\phi(w)/2}\sim (z-w)^{1/4} \,I
\ee
(where as usual $I$ stands for the identity field). A somewhat more
complicated case would be
\be
    \tau_{3}(z)\tau_{-2}(w)=(:\d\xi\d^2\xi:e^{5i\phi/2})(z)(:\eta\d\eta
    :e^{-5i\phi/2})(w)\sim(z-w)^{1/4}\, I
\ee
More generally, we get
$\tau_{2m+1}\tau_{-2n}\sim(J^\pm)^{|n-m|}I$.
These computations confirm the following fusion rule
\be
    D^+_{-1/4}\times D_{-1/4}^- = D_0
\ee
This result agrees with the second fusion in (\ref{ayfu}).
Similar calculations confirm the first and the third cases in
(\ref{ayfu}). Notice that
if the Verlinde fusions (\ref{verli}) were blind to the specification of the
representations, the corresponding fusions would be invalidated.

All fusions computed so far have been of the type $D^+\times D^-$ (since
$D_{1/2}$ can be viewed either as a highest- or a lowest-weight
representation). Consider now fusions of the
form $D^\pm\times D^\pm$. These are precisely those for which the
Verlinde formula is supposed to apply.
The simplest case of a twist product of the type $D^-\times D^-$ is
\be
    \tau_0(z)\tau_0(w)= e^{-i\phi(z)/2}e^{-i\phi(w)/2}\sim (z-w)^{-1/4}
    e^{-i\phi(w)}
\ee
It corresponds to the fusion of the ``bottom'' fields in
$D^-_{-1/4}\times D_{-1/4}^-$. The
result, however, is {\it not a twist field}.  In other words, there is no
value of $\la$ for which $e^{-i\phi}$ can be written in the form $\sigma_\la
e^{i(\la-1/2)\phi}$. This is clear since $e^{-i\phi}$ has
dimension $-1/2$ while twist fields have dimension $-1/8$.
Moreover, its products with the ghost fields
\be
    \beta(z)e^{-i\phi(w)}\sim (z-w)^{-1}(\eta e^{-2i\phi})(w),
    \qquad \gamma(z)e^{-i\phi(w)}\sim (z-w)\d\xi(w)
\ee
do not have the monodromy properties that characterize the twist fields.
Products of generic descendants in $D^-_{-1/4}\times D_{-1/4}^-$,
namely $\tau_{-2n}\times\tau_{-2m}$, yield
$J_0^+$ descendant of $e^{-i \phi}$. Therefore, the set of fields appearing
in $D^-_{-1/4}\times D_{-1/4}^-$ is $\{ (J_0^+)^ne^{-i \phi}
|n\in\ZZ_\geq\}$. They all have dimension $-1/2$.

Consider a sample product associated to $D^+_{-1/4}\times D_{-1/4}^+$:
\be
    \tau_1(z)\tau_1(w)= e^{i\phi(z)/2}e^{i\phi(w)/2}\sim (z-w)^{-1/4}
    e^{i\phi(w) }
\ee
Again, $e^{i\phi}$ is not a twist field. It is another new field,
also with dimension $-1/2$. The various fields occuring in
$D^+_{-1/4}\times D_{-1/4}^+$ are $\{ (J_0^-)^ne^{i \phi}|n\in\ZZ_\geq\}$.

Next, consider the simplest product within $D^-_{-3/4}\times D_{-3/4}^-$:
\be
    \tau_{-1}(z)\tau_{-1}(w)= (\eta e^{-3i\phi/2})(z)(\eta
    e^{-3i\phi/2})(w)\sim (z-w)^{-5/4}
    (:\d\eta\eta: e^{-3i\phi})(w)
\ee
The field appearing on the right hand side has dimension $-3/2$. It
is not a new field, however, since it can be expressed as a $J^+$
descendant of one of the new fields already found. It is easily
recognized as $J^+_1e^{-i\phi}$:
\begin{eqnarray}
       J_{1}^{+}e^{-i\phi(w)}&=&{1\over 2\pi i}\oint dz (z-w)
  J^{+}(z)e^{-i\phi(w)}\nonumber\\
       &=&{1\over 2\pi i}\oint dz {1\over 2}{1\over (z-w)}
  (e^{-3i\phi}:\partial\eta\eta:)(w)\nonumber\\
       &=&{1\over 2} (e^{-3i\phi}:\partial\eta\eta:)(w)
\end{eqnarray}
This also shows that $e^{-i\phi}$ is not associated to an
affine highest-weight state, even though it {\em is} a Virasoro
highest weight. The $su(2)$ descendants in $D^-_{-3/4}\times
D_{-3/4}^-$ are $\{(J_0^+)^n(J^+_1e^{-i\phi})|n\in\ZZ_\geq\}$.

Finally, the structure of the  product $D^+_{-3/4}\times D_{-3/4}^+$
is fixed by that of its top fields:
\be
   \tau_{2}(z)\tau_{2}(w)= (\d\xi e^{3i\phi/2})(z)(\d\xi
    e^{3i\phi/2})(w)\sim (z-w)^{-5/4}
    (:\d^2\xi\d\xi: e^{3i\phi})(w) \sim (z-w)^{-5/4}
    (J^-_1 e^{i\phi})(w)
\ee
with the descendants comprising the set
$\{(J_0^-)^n(J^-_1e^{i\phi})|n\in\ZZ_\geq\}$.

Next we consider the two remaining products $D^-_{-1/4}\times
D_{-3/4}^-$ and $D^+_{-1/4}\times D_{-3/4}^+$.
Of the first kind, we have the fusion
\be
    \tau_{0}(z)\tau_{-1}(w)=  e^{-i\phi/2}(z)\eta
    e^{-3i\phi/2}(w)\sim (z-w)^{-3/4}
    \eta e^{-2i\phi}(w)\sim  (z-w)^{-3/4}(\beta_{\frac{1}{2}}
e^{-i\phi})(w)
\ee
with descendants $\{(J^+_0)^n(\beta_{\frac{1}{2}}
e^{-i\phi})|n\in\ZZ_\geq\}$.  Associated to the second product, we have
\be
    \tau_{1}(z)\tau_{2}(w)=  e^{i\phi(z)/2}(\d\xi e^{3i\phi/2})(w)
    \sim (z-w)^{-3/4}(\d\xi e^{2i\phi})(w)\sim (z-w)^{-3/4}
    (\gamma_{\frac{1}{2}} e^{i\phi})(w)
\ee
with descendants $\{(J^-_0)^n(\gamma_{\frac{1}{2}}
e^{i\phi})|n\in\ZZ_\geq\}$.

In summary, by considering products of the form $D^\pm_{-l/4}\times
D_{-l'/4}^\pm$, $l,l'=1,3$, we
have found the following set of new fields:
\bea
\label{lili}
   && e^{-i\phi},\quad  (\beta_{\frac{1}{2}} e^{-i\phi}),\quad  (J^+_1
   e^{-i\phi}), \quad {\rm
   and~their~}J^+_0{\rm ~descendants}  \nn
   && \phantom{-}e^{i\phi},\quad\, (\gamma_{\frac{1}{2}} e^{ i\phi}),\quad
\; (J^-_1
   e^{i\phi}),
   \qquad {\rm and~their~}J^-_0{\rm ~descendants}
\eea
This provides an explicit construction of all the fields in the products
$D^\pm_{-l/4}\times D_{-l'/4}^\pm$.

Now, by taking products of twist fields with these new fields, we
produce still new fields. The simplest case is $\tau_0\times e^{-i\phi}$
which generates $e^{-3i\phi/2}$, with
dimension $-9/8$. Its product with $\tau_0$ in turn generates
$e^{-2i\phi}$, with dimension $-2$, and so on. Proceeding in this
way, we obviously get new fields at every step, with
conformal dimensions that become more and more negative.  A sample new
field occuring at the $n$-th
step is $e^{-ni\phi/2}$, of dimension $-n^2 /8$.

These computations call for a description of these new fields in
terms of the $\bg$ system\footnote{We note that a possible
way to get rid of this problem would be to show that the
$\beta\gamma$ system is a reduction of the $\eta\xi\phi$ system,
and that there are fields in the latter
that are not present in the former. This applies, in particular,
to the new fields generated in the fusion of $D_{-1/4}^\pm
\times D_{-1/4}^\pm$ as represented in the $\eta\xi\phi$ system. We
have, however, found
no indication that such a reduction is possible nor necessary.}
and the $\su(2)_{-1/2}$ admissible representations. Both points are
addressed in turn in the following two subsections.
We stress that these results immediately invalidate
the Verlinde fusion rules in the form
(\ref{verli}), and demonstrate the incompleteness of the Awata-Yamada fusion
rules (\ref{ayfu}).

\subsection{Deeper-twist fields}

The usual twist fields $\tau$ are primary fields of the
$\bg$ chiral algebra in the R sector. Apart from the vacuum, these are the
only affine primary fields in the NS sector.
However, the field structure of the $\bg$ system is much richer
than that. It contains composites of the twist fields, which will be
called {\it deeper twists}. There
is an infinite number of them, parameterized by a positive integer
$n$.  The deeper twists will be denoted
$\tau^{(n)}$. They  can be defined from their monodromy property with
respect to the ghost fields, by
demanding, for instance, that the leading terms in the OPEs be:
\be
    \beta(z)\tau^{(n)}(w)\ \propto\ (z-w)^{-n/2},\ \ \ \
    \gamma(z)\tau^{(n)}(w)\ \propto\ (z-w)^{-n/2}
\ee
In this notation, the $\tau$ introduced previously would be
$\tau^{(1)}$.

To illustrate, let us discuss $\tau^{(2)}$ in more details. We can define
the Green function $g_2^{(2)}(z,w)$ as before, i.e.,
\be
    g_2^{(2)}(z,w)={\langle
\beta(z)\gamma(w)\tau^{(2)}(2)\tau^{(2)}(1)\rangle
     \over \langle \tau^{(2)}(2)\tau^{(2)}(1)\rangle}
\ee
and find that
\be
    g_2^{(2)}(z,w)=z^{-1}w^{-1} {Az^2+(1-A)w^2\over z-w}
\ee
{}From this it follows that the dimension of $\tau^{(2)}$ is
$h=-{1\over 2}$, while its charge is $J_0^3=-{1\over 2}(2A-1)$.
In terms of the $\eta\xi$ system, one can represent it as
\be
\label{deepdeux}
    \tau^{(2)}=\sigma^{(2)}_\lambda e^{i(\lambda-1)\phi}
\ee
(so that $A={\lambda\over 2}$), with the monodromies
\be
    \eta(z)\sigma^{(2)}_\lambda(w)\sim (z-w)^{-\lambda},\ \ \ \
     \partial \xi(z)\sigma^{(2)}_\lambda(w)\sim (z-w)^{\lambda-2}
\ee
These fields $\sigma^{(2)}_\lambda$ have weight
$h=-{\lambda(2-\lambda)\over 2}$. Note that
$\sigma^{(2)}$ is {\em not} what we so far have called
a twist field in the $\eta\xi$ system because of the difference
between its expansions with $\eta$ and $\xi$.  As before, the above
relations will not always hold when $A=0$ or $A=1$.

It is important to realize that the deeper-twist fields are not new
objects.  In fact, $\tau^{(2)}_{2\lambda}$ appears naturally
in the OPE of $\tau^{(1)}_\lambda$ with itself. This can be seen most
clearly if one considers
\be
   g_4(z,w)={\langle
     \eta(z)\partial\xi(w)\tau_{1-\lambda}(4)\tau_\lambda(3)
    \tau_{1-\lambda}(2)\tau_\lambda(1)\rangle\over \langle
     \tau_{1-\lambda}(4)\tau_\lambda(3)
   \tau_{1-\lambda}(2)\tau_\lambda(1)\rangle}
\ee
Introducing the forms
\begin{eqnarray}
   \omega_1(z)&=&
    [(z-z_1)(z-z_3)]^{-\lambda}[(z-z_2)(z-z_4)]^{-1+\lambda}\nonumber\\
    \omega_2(w)&=&[(w-z_1)(w-z_3)]^{\lambda}[(w-z_2)(w-z_4)]^{1-\lambda}
\end{eqnarray}
one finds that
\begin{eqnarray}
       g_4(z,w)&=&\omega_1(z)\omega_2(w)\left[\lambda
       {(z-z_{1})(z-z_{3})(w-z_{2})(w-z_{4})\over (z-w)}\right.\nn
    &+&\left.(1-\lambda){(z-z_{2})(z-z_{4})(w-z_{1})(w-z_{3})\over (z-w)}+
    C\right]
       \end{eqnarray}
Letting $z_1\rightarrow z_3$ and $z_2\rightarrow z_4$ shows the
appearance of singularities $(z-z_1)^{-2\lambda}$ and
$(w-z_2)^{2\lambda-2}$, characteristics of the
$\sigma_{2\lambda}^{(2)}$ core of the field $\tau_{2\lambda}^{(2)}$.
The reason why the core field $\sigma_{2\lambda}^{(2)}$
(in the $c=-2$ $\eta\xi$ system) was not noticed before, is that the twist
fields that were considered had
$\lambda = \frac{1}{2}$, i.e. $\sigma_{1/2}^{(1)}$.
The point is that
the case $\lambda={1\over 2}$ is special, and the coupling
to the field $\sigma^{(2)}_{1}$ vanishes. (It vanishes both in the
numerator and the denominator of $g$, which is why it still appears
formally in the previous equations. But one can check that the OPE
coefficient vanishes at that point.) In the
$\beta\gamma$ system, twist fields generically involve
$\tau^{(2)}_\lambda$ with $\lambda$
arbitrary, and therefore no truncation occurs. This can also be seen
directly at the level of the four-point twist
correlator. From that perspective, it is the $\eta\xi$ system that appears
special, at least as long as one restricts to rational twists,
i.e., $\la\in \QQ$. The significance of an $\eta\xi$
system with irrational twists remains to be explored.

The representation (\ref{deepdeux}) generalizes to deeper twists with $n>2$:
\be \label{deprep}
    \tau^{(n)}_\lambda= \sigma^{(n)}_\lambda e^{i(\lambda-{n\over 2})\phi}
\ee
The dimension of $\sigma^{(n)}_\lambda$ is
$h=-{\lambda(n-\lambda)\over 2}$, while that of $\tau^{(n)}_\lambda$ reads
\be
    h_{\tau^{(n)}_\lambda}=-{n^2\over 8}
\ee
Thus, the spectrum of the deeper twists is {\it unbounded from below}.

Now, as one can naturally expect, the deeper twists $\tau^{(2)}$ are
exactly the fields identified in the products
$D^\pm_{-l/4}\times D_{-l'/4}^\pm$ (with $l,l'=1,3$). In
particular, we have $\tau_0^{(2)}= e^{-i\phi}$, and this
identification is further supported by the OPEs
\be
   \beta(z)e^{-i\phi}(w)\propto (z-w)^{-1},\qquad
    \gamma(z)e^{-i\phi}(w)\propto (z-w)
\ee
Similarly, we have $\tau_2^{(2)}= e^{i\phi}$. Both expressions are
indeed of the form (\ref{deprep}) with $\sigma^{(2)}_0=\sigma^{(2)}_2= I$.
It should be noticed that no deeper twists $\tau_\la^{(2)}$ for $\la$ odd
are needed to close the operator algebra in the $\beta\gamma$ system.
If they can actually be constructed, the simple presence of
$\sigma_{1/2}^{(1)}$ in the $c=-2$ sector is not sufficient because of
the zero occuring in the OPE, cf. the comment made above.
To settle whether there is a consistent theory with a closed operator
algebra containing these twists, is beyond the scope of this paper.

The images of the deeper-twist fields under the zero-mode
algebra are also deeper-twist fields. For instance,
\be
   J_0^+ e^{-i\phi}\propto
    \left(-2i\partial\phi\partial\eta\eta+\partial^2\eta\eta\right)e^{-3i\phi}
\ee
The term in bracket conspires to exactly produce the same singularities
in the expansion with $\beta$ and $\gamma$ as the initial deeper twist
$e^{-i\phi}$.

Twist fields deeper than $\tau^{(2)}$ will appear in  products
$\tau_\la(z)\tau_{\la'}^{(2)}(w)$ and more generally, in products
$\tau_\la^{(n)}(z)\tau_{\la'}^{(m)}(w)$, as long as
sign($\la$) = sign($\la'$).  We have already
identified an explicit and simple example of arbitrary deeper twist
appearing in the repeated product of
$\tau_0$ with itself, namely $\tau_0^{(n)}= e^{-in\phi/2}$.
Another simple example is $\tau_n^{(n)}= e^{in\phi/2}$, which is
similarly obtained from the repeated product of
$\tau_1$ with itself.

That settles the question of these new fields appearing in the fusions of
the type $D^\pm\times D^\pm$ from the point of view of the $\bg$ system.
We now have to see
how these deeper twists can be described from the $\su(2)$ point of view.

\subsection{Spectral flow of the $\su(2)_{-1/2}$ admissible representations}

At first sight, the identification of the deeper twists in terms of
representations of the $\su(2)_{-1/2}$ WZW model appears to be rather
problematic. Indeed, the irreducible
representations that are modular covariant are precisely the
admissible representations:
there are only four of them and their spectrum is bounded from below by
the value $-1/8$.  The question of how to generate an
unbounded spectrum from these admissible representations still has to
be addressed.

The resolution to this problem lies in a remarkable feature of the admissible
representations. Namely, they are {\em not} mapped onto themselves
under twisting, or in the $\su(2)$ terminology, under the spectral flow.
To see this, we first discuss the spectral flow which is
a symmetry transformation of the $\su(2)$ algebra:
\be
\tilde{J}_n^3 = J_n^3 - \frac{k}{2} w \delta_{n,0}, \qquad
\tilde{J}_n^{\pm}=J_{n\mp w}^{\pm}
\end{equation}
In terms of the currents themselves, these transformations read
\be
\tilde{J}^3 = {J}^3 -\frac{k}{2} w,\qquad\tilde{J}^\pm = J^\pm e^{\mp i w z}
\ee
The new Sugawara stress-energy tensor is\footnote{This result is
easily obtained starting from the classical version of the Sugawara
stress-energy tensor, $T=(1/2k)(2J^3J^3+ J^+J^-+J^-J^+)$ which becomes,
under the transformation of the currents, ${\tilde T}= (1/2k)(2J^3J^3+
J^+J^-+J^-J^+)-wJ^3+(k^2/4)w^2$. Upon quantization, the prefactor in front
of the bilinear term gets renormalized in the usual way: $k\rightarrow k+2$.}
\be
{\tilde T}= T-wJ^3+{k\over 4} w^2
\ee
or in terms of modes
\be \label{virflo}
\tilde{L}_n=L_{n}-wJ_{n}^{3}+{k\over 4}w^{2}\delta_{n,0}
\ee
In the following, we mainly need the zero-mode transformations, which
(for $k=-1/2$) read\footnote{From these transformations, we can easily
check that in spite of the quadratic character of the $L_0$
transformation, flowing successively with $w$ and $w'$ is equivalent to
flowing with $w+w'$.}
\begin{equation}
\label{flodi}
       \tilde{L}_0=L_{0}-wJ_{0}^{3}-{w^2\over 8}\; ,\qquad \tilde{J}_0^3
    = J_0^3 +{w\over 4}
\end{equation}

The spectral flow (with $w\in\ZZ$) is nothing but the action of the 
automorphisms $\pi_w$, acting on operators as
\be
\pi_w (J^\pm_m) \pi_w^{-1}= \tilde{J}^{\pm}_{m}= J^\pm_{m\mp w},
\qquad \pi_w ( J^3_m)
\pi_w^{-1}=\tilde{J^{3}_{m}}= J^3_m-{k\over 2} w\delta_{m,0}
\ee
With these relations, the action of $\pi_w$ can be
studied at the level of
the representations.\footnote{In terms of WZW models, the spectral
flow has a simple
interpretation in the classical limit. Indeed, recall that solutions to the
equations of motion have the form $g=g(z)\bar{g}(\bar{z})$. If we put
the theory on a cylinder with space period $\omega_1$, periodicity
simply requires that $g(z+\omega_1)=g(z)M$ and
$\bar{g}(\bar{z}+\omega_1)=M^{-1}g(z)$, $M$ an element of $SU(2)$ (or
$SL(2,\RR)$). The spectral flow affects such a solution by multiplying
it  by group elements whose Euler angles are linear in $z$. This
corresponds to the action of an element of the loop group which is not
continuously connected to the identity. Since $\Pi_1(SU(2))$ is trivial, this
interpretation requires the model to be defined in terms of the
$SL(2,\RR)$ group symmetry. But recall that this interpretation is action
dependent, and the existence of an action is a feature that is lost in the
$\su(2)$ case at fractional level.}

For integer level $k$, the automorphisms with
$w$ even are nothing but the inner automorphisms, related to
affine Weyl reflections. In particular, the highest-weight state is
mapped to a given
state in its affine Weyl orbit, thus to a state that lies within the
representation. Clearly,  the  application of $\pi_{\pm2w}$ ($w>0$)
on a affine highest-weight state (i.e.,  a state annihilated by
all $J^a_{n>0}$) always leads to a state
that does not have the  affine highest-weight property (although it is
still a Virasoro highest weight, which is easily  checked using
(\ref{virflo})). The remaining transformations map representations into
each others. In particular,
\be
    \pi_{-1} :\ \  j \rightarrow \frac{k}{2}-j
\ee
Therefore, the spectral flow does not generate anything new when
$k\in\NN$: the integrable
representations are simply mapped onto themselves.

The situation is quite different for $k=-1/2$. To see this immediately,
simply consider the flowed form of the vacuum state. Its flowed  dimension
(cf. (\ref{flodi})) is
$-w^2/8$, and this can be made as negative as desired.  This is a
neat signal that the
initial admissible representations are not recovered under spectral flow.

Let us consider in detail the action of $\pi_{-1}$ on the different 
admissible representations viewed as highest-weight representations.
Let us start with the vacuum representation  $D_0$,
whose highest-weight state $|0\R$ satisfies
\be
    J^3_0|0\R = 0, \qquad  J^-_{1}|0\R = 0, \qquad J^-_0|0\R = 0, \qquad
    J^+_0|0\R = 0
\ee
These relations are transformed into
\begin{eqnarray}
    &&(J^3_0-1/4)|\pi_{-1}(0)\R = 0, \qquad J^-_{0}|\pi_{-1}(0)\R = 0\nn
    &&J^-_{-1}|\pi_{-1}(0)\R = 0, \qquad  J^+_{1}|\pi_{-1}(0)\R = 0
\end{eqnarray}
This fixes $|\pi_{-1}(0)\R$ to be a lowest-weight state with
$m=-j=1/4$, so that
\be
    \pi_{-1}(D_0)= D^-_{-1/4}
\ee
Proceeding in a similar way, still focusing on the highest-weight
state, we find that
\be
   \pi_{-1}(D_{1/2})= D^-_{-3/4}\,,\qquad
   \pi_{-1}(D^+_{-1/4})= D_{0}\,, \qquad
   \pi_{-1}(D^+_{-3/4})= D^-_{1/2}
\ee
where the last relation indicates that the highest-weight state $|-3/4,-3/4\R$
is mapped to the lowest-weight state of $D_{1/2}$: $|1/2,-1/2\R$.
Understanding the labeling $\pm$ for $j=0,1/2$ in that sense, we thus have 
\be
    \pi_{-1}(D^+_{j})= D^-_{k/2-j}
\ee
Similarly, we find\footnote{To further illustrate these computations,
note that in the analysis of
$\pi_{1}(D^-_{1/2})= D^+_{-3/4}$, we find in particular that the
simplest singular vector of $D^-_{1/2}$, namely
$(J^+_0)^2|1/2,-1/2\R$, is sent to $(J^+_{-1})^2|-3/4,-3/4\R$, as it should.}
\be
   \pi_{1}(D^-_{0})= D^+_{-1/4}\,,\qquad
   \pi_{1}(D^-_{1/2})= D^+_{-3/4}\,,\qquad
   \pi_{1}(D^-_{-3/4})= D^+_{1/2}\,, \qquad
   \pi_{1}(D^-_{-1/4})= D^-_{0}
\ee
that is,
\be
    \pi_{1}(D^-_{j})= D^+_{k/2-j}
\ee
All actions of $\pi_w$ not listed above lead to representations that
are {\em not affine highest weights}. An example is given in Figure 1 with
the filled dots indicating the extremal points of the representations.
\begin{figure}
\centerline{\epsfig{figure=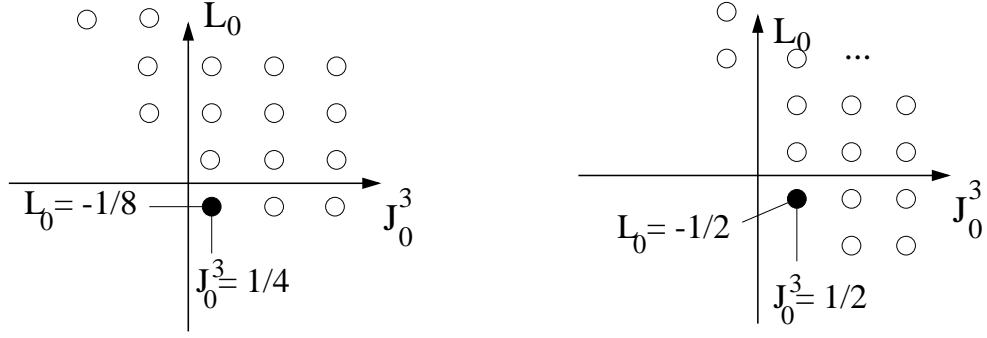,width=13cm}}
\vspace{0.3cm}
\caption{Diagram of the lowest-weight representation $\pi_{-1}(D_0)$,
   and its twist $\pi_{-2}(D_0)$. In the latter case, $L_{0}$ eigenvalues are not bounded from below.}
\label{Fig.1}
\end{figure}

Let us make this somewhat more concrete by considering the
transformation of the admissible characters.
Since the characters are defined as
\be
    \chi={\rm Tr}\, e^{2i\pi\tau(L_{0}-c/24)}e^{4i\pi z J_{0}^{3}}
\ee
they transform as follows under the spectral flow:
\begin{equation}
     \chi_j(\tau,\theta)\rightarrow \tilde{\chi}_j={\rm Tr}_{D_j}\, e^{2i\pi
\tau(\tilde{L}_{0}-c/24)}e^{4i\pi z \tilde{J}_{0}^{3}}=
e^{-i\pi\tau w^{2}/4}\,
e^{i\pi zw}\;\chi_j(\tau,z-w\tau/2)
     \end{equation}
Take the case $w=-1$. Using the transformation formulas
\begin{eqnarray}
     \theta_{1}(z+\tau/2,\tau)&=&i e^{-i\pi
z}e^{-i\pi\tau/4}\theta_{4}(z,\tau)\nonumber\\
      \theta_{2}(z+\tau/2,\tau)&=& e^{-i\pi
z}e^{-i\pi\tau/4}\theta_{3}(z,\tau)\nonumber\\
      \theta_{3}(z+\tau/2,\tau)&=& e^{-i\pi
z}e^{-i\pi\tau/4}\theta_{2}(z,\tau)\nonumber\\
        \theta_{4}(z+\tau/2,\tau)
    &=&i e^{-i\pi z}e^{-i\pi\tau/4}\theta_{1}(z,\tau)
        \end{eqnarray}
and the theta-function form (\ref{adtheta}) of the admissible characters,
we find the following transformations:
\begin{eqnarray}
        \chi_{0}&\rightarrow& {\phantom{-}}\chi_{-1/4}^{+}\nonumber\\
        \chi_{1/2}&\rightarrow& {\phantom{-}}\chi_{-3/4}^{+}\nonumber\\
        \chi_{-1/4}^{+}&\rightarrow& -\chi_{1/2}\nonumber\\
        \chi_{-3/4}^{+}&\rightarrow& {\phantom{-}}\chi_{0}\label{specfl}
        \end{eqnarray}
In view of (\ref{conj}), they are compatible with the results just
obtained. Notice the required period 4 in these maps.
Since the spectral flow shifts conformal weights
by $w^{2}/8$, an increase of $w$ by a multiple of 4 ensures
that the  weights differ by integers.

Let us now turn to fusion rules and see how the flowed
representations could appear. The key step is the assumption \cite{Gab}
that the fusion rules should be invariant under the full action of $\pi_w$:
\be
\label{gabhyp}
    \pi_w(\phi) \times \pi_{w'}(\phi') = \pi_{w+w'}(\phi \times \phi')
\ee
and not just with respect to the action of the
outer automorphism $a$ on spins (cf. (\ref{auto})).
Applying this to a simple example, we find that the product
$D_{-1/4}^-$ with itself, for example, amounts to 
\be
    D_{-1/4}^-\times D_{-1/4}^- = \pi_{-1}(D_0) \times \pi_{-1}(D_0) =
    \pi_{-2}(D_0 \times D_0) = \pi_{-2}(D_0)
\ee
The $\pi_{-2}$ flowed dimension of the vacuum state being $-1/2$, we
recover our deeper twist $\tau^{(2)}_0$.
   It is now viewed as the
   ``lowest weight'' of a representation
   that is not itself affine
   lowest-weight (cf. Fig 1).

If we consider the list of novel fields that appear in products
$D^\pm\times D^\pm$, we see that
we only need to account for the presence of $e^{\pm i\phi}$, the other
``new fields'' being natural composites with either the ghosts
or the mode currents. The two fields $e^{\pm i\phi}$ have the same
conformal dimension but they differ by their $J_0^3$ eigenvalues.  It
is thus clear that if one of the fields
corresponds to $\pi_{-2}(I)$, the other one
is $\pi_{2}(I)$. More generally, the twist fields $\tau_0^{(n)}$ and
$\tau_n^{(n)}$ are the flowed versions
$\pi_{\pm n}(I)$.\footnote{Notice that the action of the automorphism
on representations is opposite to
that on the operators and hence on their eigenvalues.}

We thus conclude that those  operators generated under
fusions computed by the $\eta\xi\phi$ representation and having
increasingly negative
dimensions, correspond to the spectrally flowed representations. In
other words, our free-field
computations corroborates the assumption (\ref{gabhyp}).

In light of these observations, it is natural to expect that the 
operator algebra only closes if all flowed representations
are included.
Since both highest- and lowest-weight representations are generated  
for $j=-1/4, -3/4$ by acting
with $\pi_{\pm 1}$ on $j=0,1/2$, none of the previous
interpretations of the diagonal modular invariant can be correct.
Some fields were missing in each case.
We revisit the invariant in the next
section and show how a consistent theory can be constructed.
We find that the modular invariant is obtained essentially by summing over
the orbits of the spectral flow.

\subsection{The $\su(2)_{-1/2}$ partition function revisited}

Let us look again at the character of an admissible highest-weight
representation:
\be
   \chi_{j}^+(q,y) = {\rm Tr}_{D_j^+} q^{L_0+1/24}\,  y^{2J^3_0}
    \qquad ( q=e^{2 \pi i\tau}\,,\quad y= e^{2\pi i z})
\label{ch+}
\ee
We suppose that $|q|<1$ and focus on the expansion of
the character in terms of the variable
$y$. As a function of $y$, this character has poles, which means that
the summed expression (\ref{ch+})
is only defined in a particular region in the complex plane.
In other words, the character of the spin-$j$ representation is given
by the function
{\it and} the specification of a region of convergence.  In this
case, the region is given by the annulus $1<|y|<1/|q|^{1/2}$.

Once we flow the representation, the new character converges in a region
determined by the flow. Since the character transforms under the flow as
\be
    \chi_{j,w}^+(q,y) = q^{- w^2/8} y^{w/2} \chi_{j}^+(q,yq^{-w/2})
\label{chi2}
\ee
the new region of convergence becomes (cf. \cite{Fei})
\be
    |q|^{w/2} < |y| < |q|^{(w-1)/2}
\ee
Despite the notation, (\ref{chi2}) may result in a character associated
to a lowest-weight representation (cf. (\ref{conj})). 

Now we know that the functional form above falls back on one of our
original functions, but we should be careful. Since the original region of
convergence is mapped to a new region of convergence  for the
``flowed character'' (when interpreting
the character in terms of an operator or representation), we get
the following dimension for the flowed operator:
\be
    h_{j,w}^+ = \frac{j(j+1)}{3/2}-w j -\frac{1}{8}w^2
\ee
(where the upper index $+$ indicates that we  use $m=j$ here).
For instance, for $j=-1/4$, we see that $w=1$ falls back on the
identity, while $w=-1$ gives $h= -1/2$ which corresponds to
$\pi_{-2}(I)$. In this way we see all the fields, but they are linked to
particular convergence regions. In particular, the highest- and
lowest-weight representations are
all there. It is only that, say, $D_{-1/4}^-$ and $D_{-3/4}^+$ have
different regions of convergence. Stated differently, flowing among
representations amounts to  perform {\em analytic continuations}.

The fact that we have to define convergence regions
is not obvious from the original ``trace point of view''.
But since we know the result of the sum (cf. (\ref{cara})), we see that
the reason is due to the singularities in the summed
function. The presence of these singularities is a
feature particular to the fractional-level case. For $k$ integer, the sum is a
{\em holomorphic} character function, converging
everywhere in the plane. As a result, performing a spectral flow
does not change the region of convergence. As already mentioned, it
is a characteristic of the integer-level case that the flowed integrable
representations are all mapped back to integrable representations.

Let us stress the following. Even though the set of admissible character
functions \cite{Fei} (when combined to form the Kac-Wakimoto invariant)
is invariant under the modular transformations, the physical states are
associated with particular expansions, in well-defined regions of
convergence. Hence, a character function is not mapped to a unique
field or $\su(2)$ module. Rather, it is the decomposition on states
which is required in the determination of the spectrum.

Keeping this in mind, we can revisit the modular invariance. When a
transformation of the form $\tau\rightarrow \tau+1$ is performed,
character functions are indeed mapped into each other. The
associated regions of convergence, on the other hand, are {\em not}
mapped into each other. They are mapped into the regions of convergence
of the spectrally flowed operators.
This means that to have a full-fledged modular invariant, one
has to include the infinite set of flowed representations. The invariant
should really be interpreted as a sum over all regions of convergence
of all the expansion series. Thus, the partition
function that includes all the twisted modules reads
\be
    Z\ =\ \sum_{w\in \ZZ}\ \sum_{j=0,1/2}|q^{-w^2/8}y^{w/2}
\chi_j(q,yq^{-w/2})|^2
\ee
Since the partition functions have period four under the flow, we
finally arrive at the usual functional form
(here the superscript is included to distinguish the functions,
not the representations)
\be
 Z\ =\ \sum_{{\cal D}}\left(|\chi_0^+(q,y)|^2 +
 |\chi_{1/2}^+(q,y)|^2 + |\chi_{-1/4}^+(q,y)|^2 + |\chi_{-3/4}^+(q,y)|^2
  \right)
\ee
This sum is understood to be over all domains of convergences over which
we expand the functions to get the characters. 
Thus functionally, there is an infinite constant
multiplying the usual partition function.

Now, with these comments, one should be careful when interpreting
the Verlinde formula. The potential problem linked to
the fact that the modular transformations
relate different regions of convergences, is not taken into account
in the derivation of the Verlinde formula. Therefore, it
is well established only for integrable representations, or equivalently,
for holomorphic character functions. As we have seen in our case,
twisted modules appear under the modular transformations, meaning that
we do not flow back onto the original set of fields when going around
cycles on the torus. This clearly indicates that the
Verlinde formula does not apply to the $\su(2)_{-1/2}$ WZW model
(and nor to more general fractional-level WZW models). The previous belief
that the four admissible fields close under fusion is incorrect,
even though it is naively (but {\em only} naively) supported by the
Verlinde formula.

\section{Conclusion}

The main conclusion of our study is that the $\su(2)_{-1/2}$
model defined algebraically \`a la Kac and Wakimoto \cite{KK}, and the
$\beta\gamma$ system with a standard choice of normal ordering in the
R sector (i.e., the ground state is annihilated by one of the ghost 
zero modes), are not rational CFTs 
in the conventional sense.\footnote{In the terminology of
\cite{MooreS}, these are quasi-rational CFTs in that there is an infinite but
countable number of characters in the partition function, while only a
finite number of fields appear in any given fusion.}  
In both formulations of this $c=-1$ model, the spectra contain operators of
arbitrarily large negative dimensions which are not primary fields with
respect to the chiral algebra (either $\bg$ or $\su(2)_{-1/2}$).  On the WZW
side, a formal, yet meaningful theory, can be obtained by extending the basic
set of admissible fields to include their orbits  under the spectral flow.
Viewed from the $\bg$ perspective, this amounts to taking into account an
infinite number of deeper twists. We stress that without these extensions, the
theories are not consistent. This can be seen either at the level of the
fusion rules, which otherwise do not close, or at the level of
the modular invariant, since
modular transformations map characters to their flowed versions.

Let us re-phrase this conclusion somewhat in order to settle 
some loose points in our initial discussion of the $\bg$ system in 
Section 2.1. As pointed out there, the
$\beta\gamma$ system with the usual  highest-weight
conditions (\ref{hwcond}) is plagued with divergences
because the functional integral cannot be properly
defined. Naive analytic continuation leads to a
partition function which is essentially
the inverse of $\hbox{det }\Delta$. A better procedure is to define
the model in terms of the associated $\su(2)$ CFT. We have seen, however, 
that this WZW model has a rich operator content.
Its spectrum extends infinitely beyond the simple set of
admissible representations, by including all their images under the spectral
flow. Nevertheless, there is a way to define
characters of the flowed representations using analytic continuation such that
the partition function of the complicated theory with no ground state,
coincides formally with $1/\hbox{det }\Delta$ in the vicinity of $z=0$. The
naive result is thereby put into context. 
Ultimately, the original singularity at
$z=0$ appears as one copy of an infinite number of singularities, and the
positions of these singularities determine regions of convergence in the
complex plane that are associated to the characters of the various deeper
twists. We thus see that, roughly, the original singularity hides two types
of interrelated infinities: the infinite degeneracy of the twist 
fields (which are distinguished by their
$u(1)$ charges), and the infinite number of deeper twists (each of 
which being degenerate). The spectrum is unbounded from below.

It should be clear that the choice of the highest-weight
conditions (\ref{hwcond}) is not essential, since the alternative
conditions (\ref{llwcond}) are recovered by flowing (cf. the analysis of
Section 6.3 where $D^\pm$ representations are mapped into each others by the
action of $\pi_{\pm1}$). Note, however, that the flow cannot generate a
$\bg$ vacuum that is not annihilated by any of the ghost zero modes.
Such a vacuum would
lead to representations that are neither highest nor lowest weight.
They would extend infinitely in both directions (and are called 
continuous representations).

A particularity of those $\bg$ twist fields that have been found to 
be organized in infinite-dimensional 
$\su(2)_{-1/2}$ representations, is that they live entirely in the
free-fermionic sector of the $c=-2$ component. 
In other words, they do not involve those twists of the $\eta\xi$ 
sector which have no fermionic
description. This allows the theory to avoid, at least in this version,
the logarithmic extensions  found in $c=-2$ CFT.
Clearly, it is possible to introduce operators going beyond the model 
we have studied. The symplectic fermions provide an example,
where the extension amounts to including two fermionic zero modes.
Preliminary results show the presence of logarithms in that case. 
This will be the subject of our forthcoming paper \cite{LMRS}.

We stress that the
$c=-1$ model, independently of its physical applications, is a particularly
good laboratory because it admits  a faithful free-field
representation.\footnote{In that vein, notice that the representation
(\ref{Jferm}) can be viewed as a non-unitary parafermionic description of the
$\su(2)_k/ \widehat{u}(1)$ coset, where the negative
sign of the level is absorbed in a re-definition of the  boson metric.
Here the dimension of the basic parafermions (represented by $:\d\eta 
\eta:$ and $:\d^2\xi\d\xi:$) is thus 3, and the parafermionic 
algebra is bound to reduce to the triplet algebra of
\cite{KauS}.} It allows us to study general properties of non-unitary WZW 
models in a very controlled way.
Note, however, that if we measure the complexity of a non-unitary
WZW model by the number of its admissible representations, the
$\su(2)_{-1/2}$ model is not the simplest one. Indeed, it has four
admissible representations, while the model with the least possible
number of admissible representations is
$\su(2)_{-4/3}$, with three (the vacuum and two fields of
dimension $-1/3$).\footnote{Note that both models are associated to a
Virasoro $c(1,p)$ model up to a
$u(1)$ factor (with $p=1,2$, respectively) -- one of the lowest-dimensional 
fields in each case being associated to
$\phi_{1,p}$.} This latter model has central charge $c=-6$, and
has been studied extensively in \cite{Gab}. There it
is indicated that logarithms (originating from indecomposable $su(2)$
representations) may be unavoidable. This may  not be  a generic 
feature of fractional-level models, as our analysis has shown.
It is still an open question, though, to understand the origin
of the logarithms in the $\su(2)_{-4/3}$ WZW model, as it is
presented in \cite{Gab}, and to understand the differences from the
$\su(2)_{-1/2}$ model described here and in \cite{LMRS} (for its logarithmic
lift).

It should
be clear that our main conclusion concerning the non-rational character of the
WZW models at fractional level, is independent of the specific value of $k$
studied. Although the presence of an unbounded negative spectrum  can
be traced back to the sign of $k$ (cf. the expression for the flowed dimension
of the vacuum), the non-analytic property of the admissible characters is a
generic feature.

The presence of a spectrum of dimensions which is unbounded from below 
casts doubts about proposals to use $\beta\gamma$ systems in the  
CFT description of fixed points in  disordered electronic systems 
\cite{BLhall}. On the other hand,  
we have seen that some objects like the partition function of the 
$\beta\gamma$ system are, in a formal sense,
insensitive to the unboundedness of the spectrum. It is thus 
plausible that, at the same level of formality, 
useful physical quantities {\em can} be calculated using such CFT methods. 
A somewhat related situation 
occured in the study of the $U(1,1)$ WZW model
and the Alexander polynomial \cite{RozS}. We feel, however, 
that considerable care has to be exercised when employing these CFT
techniques. 

\vskip0.5cm
\noindent {\bf ACKNOWLEDGMENTS}\\[.3cm]
FL, PM and HS would like to thank IPAM for its hospitality during the
Fall workshop CFT 2001 where this work was initiated. The work of FL
and PM is supported by NSERC, that of JR by a CRM-ISM Postdoctoral 
Fellowship, and that of HS by the DOE and the NSF.

\end{document}